\documentclass[
    twocolumn
]{aastex631}

\def\lesssim{\mathrel{\hbox{\rlap{\hbox{\lower4pt\hbox{$\sim$}}}\hbox{$<$}}}}
\def\gtrsim{\mathrel{\hbox{\rlap{\hbox{\lower4pt\hbox{$\sim$}}}\hbox{$>$}}}}

\newcommand{\eg}{\mbox{e.g.}}
\newcommand{\ie}{\mbox{i.e.}}

\newcommand{\ips}{\ensuremath{i_{\textsc{ps}}}}

\newcommand{\grizy}{\ensuremath{grizy_{\textsc{ps}}}}

\newcommand{\izy}{\ensuremath{izy_{\textsc{ps}}}}

\newcommand{\msun}{\mbox{M$_{\odot}$}}

\newcommand{\rpro}{\mbox{$r$-process}}

\newcommand{\gfo}{\mbox{AT\,2017gfo}}
\newcommand{\SNxx}[1]{\mbox{SN\,#1}}
\newcommand{\ATxx}[1]{\mbox{AT\,#1}}
\newcommand{\GRBxx}[1]{\mbox{GRB\,#1}}
\newcommand{\GWxx}[1]{\mbox{GW\,#1}}

\newcommand{\black}[1]{{\textcolor{black}{#1}}}

\newcommand{\I}{{\sc i}}
\newcommand{\II}{{\sc ii}}

\newcommand{\V}{{\sc v}}

\newcommand{\X}{{\sc x}}

\newcommand{\highlighter}[1]{\black{#1}}

\def\GWsource{\highlighter{S250818k}}

\def\SNName{\highlighter{\SNxx{2025ulz}}}

\def\PSCoords{\highlighter{\mbox{${\rm RA} = 237.97584^\circ$}, \mbox{${\rm Dec} = +30.90241^\circ$}}}

\def\UTGW{\highlighter{\mbox{2025-08-18 01:20:06.030}}}    
\def\MJDGW{\highlighter{\mbox{60905.055625}}}    

\def\alertMJDGW{\highlighter{\mbox{60905.055961}}}    

\def\GWDist{\highlighter{\mbox{$237 \pm 62$\,Mpc}}}    

\def\MJDPSo{\highlighter{\mbox{60907.298193}}}   

\def\MJDdiffo{\highlighter{\mbox{2.2426}}}    

\def\whichMap{\highlighter{\texttt{bilby.fits}}}    

\def\HostGalName{\highlighter{WISEA\,J155154.15+305409.2}}          
\def\Hostz{\highlighter{$0.0849 \pm 0.0003$}}        
\def\HostDist{\highlighter{$\approx 401$\,Mpc}}     

\def\EBsubV{\highlighter{0.0244}}    

\def\Ag{\highlighter{0.090}}    
\def\Ar{\highlighter{0.065}}    
\def\Ai{\highlighter{0.048}}    
\def\Az{\highlighter{0.038}}    
\def\Ay{\highlighter{0.031}}    
\def\Aw{\highlighter{0.067}}    

\def\H0{\highlighter{$H_0 = 67.4$\,km\,s$^{-1}$\,Mpc$^{-1}$}} 

\def\ATLASforceddays{\highlighter{50}} 

\usepackage[normalem]{ulem}

\usepackage{graphicx}	
\usepackage{amsmath}	

\usepackage{orcidlink}

\usepackage{nicefrac}
\usepackage{multirow}
\usepackage{tabularx}
\usepackage[normalem]{ulem}
\usepackage{colortbl}

\usepackage{longtable}
\usepackage{booktabs}

\usepackage{xfrac}
\usepackage{array}
\usepackage{comment}
\usepackage{xcolor}

\usepackage{subfigure}

\received{\today}
\revised{TBD}
\accepted{TBD}

\shorttitle{}
\shortauthors{}

\begin{document}

\title{Pan-STARRS follow-up of the gravitational-wave event \GWsource\ and the lightcurve of \SNName}

\correspondingauthor{James~H.~Gillanders}
\email{james.gillanders@physics.ox.ac.uk}

\author[0000-0002-8094-6108]{J.~H.~Gillanders}
\affil{Astrophysics sub-Department, Department of Physics, University of Oxford, Keble Road, Oxford, OX1 3RH, UK}
\author[0000-0003-1059-9603]{M.~E.~Huber}
\affiliation{Institute for Astronomy, University of Hawai'i, 2680 Woodlawn Drive, Honolulu, HI 96822, USA}
\author[0000-0002-2555-3192]{M.~Nicholl}
\affil{Astrophysics Research Centre, School of Mathematics and Physics, Queen's University Belfast, BT7 1NN, UK}
\author[0000-0002-8229-1731]{S.~J.~Smartt}
\affil{Astrophysics sub-Department, Department of Physics, University of Oxford, Keble Road, Oxford, OX1 3RH, UK}
\affil{Astrophysics Research Centre, School of Mathematics and Physics, Queen's University Belfast, BT7 1NN, UK}
\author[0000-0001-9535-3199]{K.~W.~Smith}
\affil{Astrophysics sub-Department, Department of Physics, University of Oxford, Keble Road, Oxford, OX1 3RH, UK}
\affil{Astrophysics Research Centre, School of Mathematics and Physics, Queen's University Belfast, BT7 1NN, UK}
\author[0000-0001-6965-7789]{K.~C.~Chambers}
\affil{Institute for Astronomy, University of Hawai'i, 2680 Woodlawn Drive, Honolulu, HI 96822, USA}
\author[0000-0002-1229-2499]{D.~R.~Young} 
\affil{Astrophysics Research Centre, School of Mathematics and Physics, Queen's University Belfast, BT7 1NN, UK}
\author[0009-0004-5681-545X]{J.~W.~Tweddle} 
\affil{Astrophysics sub-Department, Department of Physics, University of Oxford, Keble Road, Oxford, OX1 3RH, UK}
\author[0000-0003-4524-6883]{S.~Srivastav}
\affil{Astrophysics sub-Department, Department of Physics, University of Oxford, Keble Road, Oxford, OX1 3RH, UK}
\author[0000-0003-1916-0664]{M.~D.~Fulton}
\affil{Astrophysics Research Centre, School of Mathematics and Physics, Queen's University Belfast, BT7 1NN, UK}
\author[0000-0002-3424-8528]{F.~Stoppa}
\affil{Astrophysics sub-Department, Department of Physics, University of Oxford, Keble Road, Oxford, OX1 3RH, UK}
\author[0000-0002-6639-6533]{G.~S.~H.~Paek}
\affiliation{Institute for Astronomy, University of Hawai'i, 2680 Woodlawn Drive, Honolulu, HI 96822, USA}
\author[0000-0002-9085-8187]{A.~Aamer}
\affil{Astrophysics Research Centre, School of Mathematics and Physics, Queen's University Belfast, BT7 1NN, UK}
\author[0000-0002-8134-2592]{M.~R.~Alarcon}
\affiliation{Instituto de Astrofísica de Canarias (IAC), C/ Vía Láctea, s/n, E-38205, La Laguna, Spain}
\affiliation{Departamento de Astrofísica, Universidad de La Laguna (ULL), E-38206 La Laguna, Canarias, Spain}
\author[0000-0003-2734-1895]{A.~Andersson}
\affil{Astrophysics sub-Department, Department of Physics, University of Oxford, Keble Road, Oxford, OX1 3RH, UK}
\author[0000-0002-9928-0369]{A.~Aryan}
\affiliation{Graduate Institute of Astronomy, National Central University, 300 Jhongda Road, 32001 Jhongli, Taiwan}
\author[0000-0002-4449-9152]{K.~Auchettl}
\affil{Department of Astronomy and Astrophysics, University of California, Santa Cruz, CA 95064, USA}
\affil{OzGrav, School of Physics, The University of Melbourne, VIC 3010, Australia}
\author[0000-0002-1066-6098]{T.-W.~Chen}
\affil{Graduate Institute of Astronomy, National Central University, 300 Jhongda Road, 32001 Jhongli, Taiwan}
\author[0000-0001-5486-2747]{T.~de~Boer}
\affil{Institute for Astronomy, University of Hawai'i, 2680 Woodlawn Drive, Honolulu, HI 96822, USA}
\author[0000-0002-5105-344X]{A.~K.~H.~Kong}
\affiliation{Institute of Astronomy, National Tsing Hua University, Hsinchu 300044, Taiwan}
\author[0000-0002-9214-337X]{J.~Licandro}
\affiliation{Instituto de Astrofísica de Canarias (IAC), C/ Vía Láctea, s/n, E-38205, La Laguna, Spain}
\affiliation{Departamento de Astrofísica, Universidad de La Laguna (ULL), E-38206 La Laguna, Canarias, Spain}
\author[0000-0002-9438-3617]{T.~Lowe}
\affiliation{Institute for Astronomy, University of Hawai'i, 2680 Woodlawn Drive, Honolulu, HI 96822, USA}
\author[0009-0000-6521-8842]{D.~Magill}
\affiliation{Astrophysics Research Centre, School of Mathematics and Physics, Queen's University Belfast, BT7 1NN, UK}
\author[0000-0002-7965-2815]{E.~A.~Magnier}
\affiliation{Institute for Astronomy, University of Hawai'i, 2680 Woodlawn Drive, Honolulu, HI 96822, USA}
\author[0009-0003-8803-8643]{P.~Minguez}
\affiliation{Institute for Astronomy, University of Hawai'i, 2680 Woodlawn Drive, Honolulu, HI 96822, USA}
\author[0000-0001-8385-3727]{T.~Moore}
\affil{Space Telescope Science Institute, 3700 San Martin Drive, Baltimore, MD 21218, USA}
\author[0000-0003-0006-0188]{G.~Pignata}
\affil{Instituto de Alta Investigaci\'on, Universidad de Tarapac\'a, Casilla 7D, Arica, Chile}
\author[0000-0002-4410-5387]{A.~Rest}
\affiliation{Space Telescope Science Institute, 3700 San Martin Drive, Baltimore, MD 21218, USA}
\affiliation{Department of Physics and Astronomy, Johns Hopkins University, Baltimore, MD 21218, USA}
\author[0000-0002-2394-0711]{M.~Serra-Ricart}
\affiliation{Light Bridges, SL.\ Observatorio Astronómico del Teide.\ Carretera del Observatorio del Teide, s/n, Güímar, \\ Santa Cruz de Tenerife, Spain}
\affiliation{Instituto de Astrofísica de Canarias (IAC), C/ Vía Láctea, s/n, E-38205, La Laguna, Spain}
\affiliation{Departamento de Astrofísica, Universidad de La Laguna (ULL), E-38206 La Laguna, Canarias, Spain}
\author[0000-0003-4631-1149]{B.~J.~Shappee}
\affiliation{Institute for Astronomy, University of Hawai'i, 2680 Woodlawn Drive, Honolulu, HI 96822, USA}
\author[0000-0001-8605-5608]{I.~A.~Smith}
\affiliation{Institute for Astronomy, University of Hawai'i, 34 Ohia Ku St., Pukalani, HI 96768-8288, USA}
\author[0000-0002-2471-8442]{M.~A.~Tucker}
\altaffiliation{CCAPP Fellow}
\affiliation{Center for Cosmology and AstroParticle Physics, 191 W Woodruff Ave, Columbus, OH 43210}
\affiliation{Department of Astronomy, The Ohio State University, 140 W 18th Ave, Columbus, OH 43210}
\author[0000-0002-1341-0952]{R.~Wainscoat}
\affiliation{Institute for Astronomy, University of Hawai'i, 2680 Woodlawn Drive, Honolulu, HI 96822, USA}

\begin{abstract}
    Kilonovae are the scientifically rich, but observationally elusive, optical transient phenomena associated with compact binary mergers. Only a handful of events have been discovered to date, all through multi-wavelength (gamma ray) and multi-messenger (gravitational wave) signals. Given their scarcity, it is important to maximise the discovery possibility of new kilonova events. To this end, we present our follow-up observations of the gravitational-wave signal, S250818k, a plausible binary neutron star merger at a distance of $237 \pm 62$\,Mpc. Pan-STARRS tiled 286 and 318 square degrees (32\% and 34\% of the 90\% sky localisation region) within 3 and 7 days of the GW signal, respectively. ATLAS covered 65\% of the skymap within 3 days, but with lower sensitivity. These observations uncovered 47 new transients; however, none were deemed to be linked to S250818k. We undertook an expansive follow-up campaign of AT\,2025ulz, the purported counterpart to S250818k. The $griz$-band lightcurve, combined with our redshift measurement ($z = 0.0849 \pm 0.0003$) all indicate that SN\,2025ulz is a SN~IIb, and thus not the counterpart to S250818k. We rule out the presence of a AT\,2017gfo-like kilonova within $\approx 27$\% of the distance posterior sampled by our Pan-STARRS pointings ($\approx 9.1$\% across the total 90\% three-dimensional sky localisation). We demonstrate that early observations are optimal for probing the distance posterior of the three-dimensional gravitational-wave skymap, and that SN\,2025ulz was a plausible kilonova candidate for $\lesssim 5$~days, before ultimately being ruled out. 
\end{abstract}


\section{Introduction} \label{SEC: Introduction}

The historic gravitational wave (GW) event \GWxx{170817}, resulting from a binary neutron star (BNS) merger \citep{Abbott2017_GW170817discovery}, produced the short gamma-ray burst (GRB)~170817A \citep{Abbott2017GRB}, and the rapidly evolving ultraviolet--optical--infrared transient \gfo, the first confirmed kilonova (KN) event \citep{MMApaper2017, Arcavi2017, Coulter2017, Lipunov17, SoaresSantos2017, Tanvir2017, Valenti2017}. Further high-cadence spectrophotometric observations followed, which confirmed the uniqueness of the EM transient \citep{Chornock2017, McCully2017, Nicholl2017, Pian2017, Shappee2017, Smartt2017}. Extensive data acquisition and modelling followed \citep{Andreoni2017, Cowperthwaite2017, Drout2017, Evans2017, Kasliwal2017, Kilpatrick2017, Tanvir2017, Troja2017, Utsumi2017}, which provided excellent matches to theory predictions, demonstrating that \gfo\ was powered by the radioactive decay of neutron-rich \rpro\ material \citep{Li1998, Metzger2010, Kasen2013, Kasen2017, Tanaka2013, Rosswog2017}.

Although other NS-bearing systems have since been detected by LIGO--Virgo--KAGRA (LVK), including the BNS merger \GWxx{190425} \citep{190425} and the neutron star -- black hole (NSBH) mergers \GWxx{190814} \citep{190814}, \GWxx{200105}, \GWxx{200115} \citep{NSBH}, \GWxx{230518} and \GWxx{230529} \citep{GW230519, GW230518_GW230519}, \GWxx{170817} has remained the only GW source with a confirmed EM counterpart.

No KNe have been uncovered by wide-field surveys (such as ZTF and Pan-STARRS; \citealt{Andreoni2020, Fulton2025}). \cite{Rastinejad2025} present analysis of eight proposed KNe (including \gfo) associated with GRB signals. Of these, the three best-sampled events are the recent GRBs 160821B \citep{Lamb2019}, 211211A \citep{Rastinejad2022, Troja2022} and 230307A \citep{Gillanders2023, Levan2024, Yang2024}. However, GRB-discovered KNe suffer from early-time contamination from the on-axis afterglow, an inevitable consequence of the fact we are only sensitive to on-axis events. GW signals are one of the best methods to ultimately enable uncontaminated early-time observations of KNe.

In this manuscript, we outline our follow-up observations of the GW event \GWsource\ -- possibly just the third astrophysically real BNS merger event -- with Pan-STARRS and ATLAS. We also summarise our targeted observational campaign for \SNName, which was proposed as a candidate optical counterpart \citep{Stein2025GCN.41414}. \GWsource\ was discovered on MJD~\MJDGW\ (\UTGW~UTC) and the alert was released at MJD~\alertMJDGW\ \citep{LVKGCNDiscovery}. This was initially classified as either a BNS merger (29\% probability) or a terrestrial source (71\% probability) by LVK, with an estimated distance of $263 \pm 75$\,Mpc. A later improvement in the analysis, presented in the \whichMap\ skymap, updated the distance estimate to \GWDist, without changing the source classification probabilities.

Throughout, we adopt Planck $\Lambda$CDM cosmology with a Hubble constant, \H0, $\Omega_{\textrm{M}} = 0.315$ and \mbox{$\Omega_{\Lambda} = 0.685$} \citep{Planck2020}. Also, all magnitudes are presented in the AB system \citep{Oke1983}.

\section{\GWsource} \label{SEC: S250818k}

\begin{figure*}
    \centering
    \includegraphics[width=0.8\linewidth]{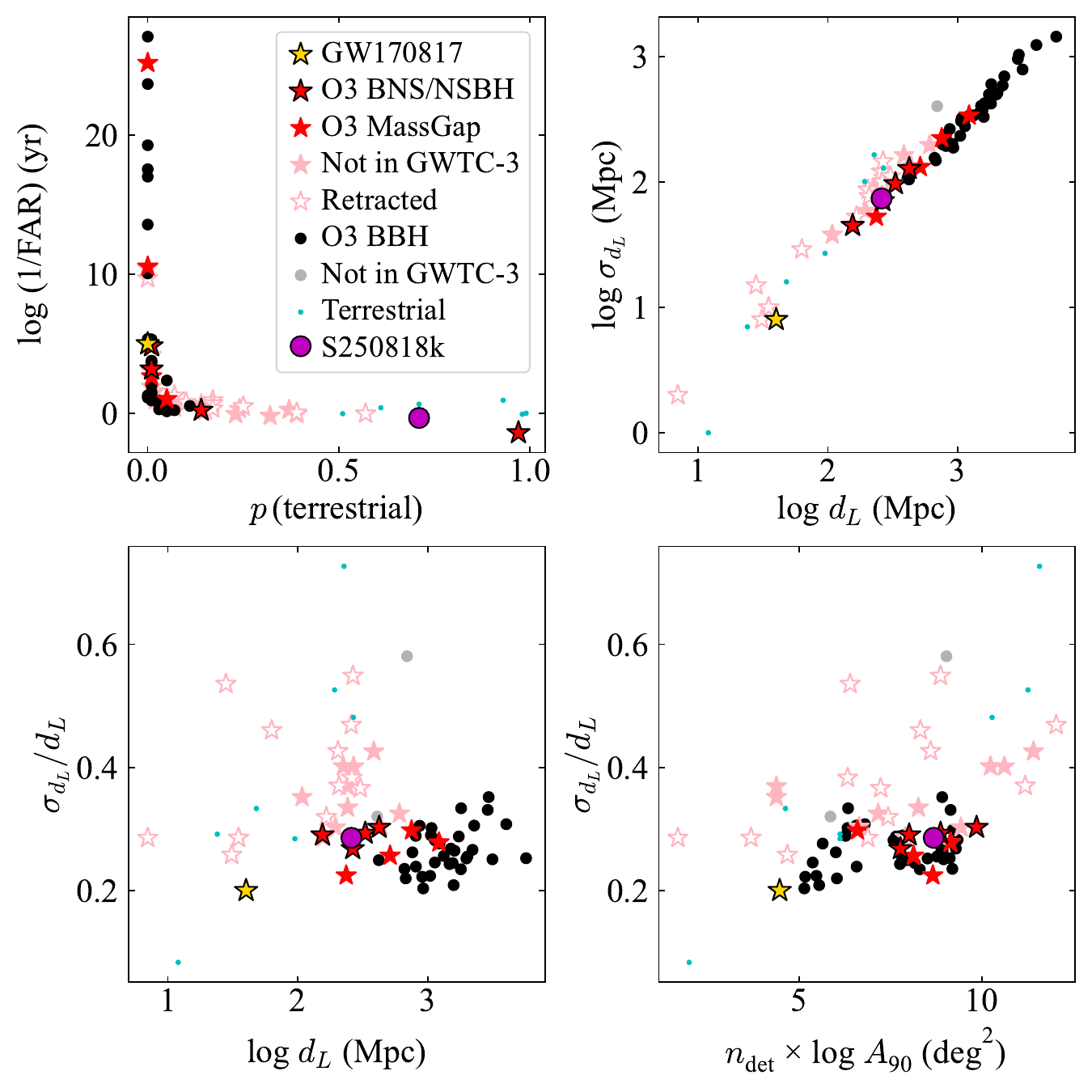}
    \caption{
        Comparison of the GW signal properties of \GWsource\ to those from the LVK GWTC-3 catalogue \citep{Abbott2023_GWTC3}, following \cite{Nicholl2025}.
    }
    \label{FIG: GW properties}
\end{figure*}

\GWsource\ was discovered on MJD~\MJDGW\ and announced publicly on MJD~\alertMJDGW\ as a sub-threshold GW event, with a false-alarm rate (FAR) of 2.1\,yr$^{-1}$ \citep{LVKGCNDiscovery}. As such, it did not pass our trigger criteria for immediate follow-up with either Pan-STARRS or ATLAS, due to its high probability of terrestrial origin ($p_{\rm terr} = 71$\%). However, additional investigation of the properties of the GW signal compared with the GW triggers from LVK's third observing run \citep[O3;][]{Abbott2023_GWTC3}, show that, despite the apparent high likelihood of terrestrial origin, the signal properties broadly match those of previously reported high-significance candidate merger events.

Specifically, we apply the method of \cite{Nicholl2025} to analyse the location of \GWsource\ in the multi-dimensional parameter space of FAR, $p_{\rm terr}$, luminosity distance ($d_L$), distance uncertainty ($\sigma_{d_L}$), and 90\% localisation area ($A_{90}$) multiplied by the number of active GW detectors ($n_{\rm det}$), as reported in low latency. The results are shown in Figure~\ref{FIG: GW properties}. \GWsource\ lies in the typical distribution for the FAR~vs.~$p_{\rm terr}$ for O3 triggers, though in a region dominated by spurious sources that were later retracted. Unsurprisingly, these parameters are correlated, with a high $p_{\rm terr}$ for events with a high FAR. However, \GWsource\ does not have the highest $p_{\rm terr}$ for a candidate NS-bearing GW signal; this title belongs to the NSBH merger S200105ae, which, despite having $p_{\rm terr} = 0.97$ (and ${\rm FAR} > 24$~year$^{-1}$) in low latency did survive in the offline analysis, and was included in the third and fourth Gravitational Wave Transient Catalogues (GWTC-3 and GWTC-4) as a marginal event \citep{Abbott2023_GWTC3, LVK2025_GWTC4}. This event was also subject to intensive EM follow-up \citep{Anand2021}, though without an identified counterpart.

In all other parameters, \GWsource\ sits in the locus of real events, away from the bulk of spurious ones. As noted by \cite{Nicholl2025}, the strongest discriminator between real and spurious GW detections arises from the $\sigma_{d_L}$/$d_L$~vs.~$d_L$ parameter space; here we find that real events tend to lie in the lower right of the figure, whereas most of the spurious detections lie in the upper left region. \GWsource\ also sits within the locus of NS-bearing events in terms of its localisation area per detector (we also note that the LVK skymap for this event conformed to the expected smooth `banana' shape; see Figure~\ref{FIG: Skymap}). Based on these considerations, it would be unsurprising if \GWsource\ was confirmed as a real astrophysical signal, and as such, we initiated our GW follow-up program with Pan-STARRS.

We do note one surprising parameter from the low latency products. The chirp mass in \GWsource\ was reported to lie (with 100\% probability) in the $0.1 - 0.87$\,\msun\ bin. This parameter was not made public in low latency during O3, so we are unable to compare this to our control population of previous GW triggers. The constituent NS masses can be derived from the chirp mass ($\mathcal{M}$) via:
\begin{equation}
    \mathcal{M} = \frac{(m_1 \cdot m_2)^{3/5}}{(m_1 + m_2)^{1/5}},
\end{equation}
where $m_1$ and $m_2$ are the heavier and lighter NS masses, respectively. Expressing the mass of the lighter NS in terms of fractional mass of the heavier NS (\ie, $m_2 = \eta \cdot m_1$, where $0 < \eta \leq 1$), and rearranging gives:
\begin{equation}
    m_1 = \left( \frac{\eta + 1}{\eta^3} \right)^{1/5} \cdot \mathcal{M}
\end{equation}
For an equal-mass merger ($\eta = 1$), the upper edge of the chirp mass bin ($\mathcal{M} = 0.87$\,\msun) corresponds to NS masses, $m_1 \equiv m_2 = 1.0$\,\msun. For all forms of unequal mergers and for all chirp masses within the $0.1 - 0.87$\,\msun\ mass bin, $m_2 < 1$\,\msun. Thus, if this source is confirmed to be astrophysically real {\it and} the low-latency chirp mass is reliable, at least one of the components must have a mass $\leq 1$\,\msun.

\section{Scanning the skymap of \GWsource} \label{SEC: Scanning the skymap of S250818k}

\subsection{Pan-STARRS} \label{SEC: Scanning the skymap of S250818k - Pan-STARRS}

Pan-STARRS (PS) is a twin 1.8-m telescope system (Pan-STARRS1 and Pan-STARRS2), both situated atop Haleakala mountain on the Hawaiian island of Maui \citep{Chambers2016arXiv_PanSTARRS1}. Pan-STARRS1 (PS1) has a 1.4~gigapixel camera and the 0.26~arcsec pixels provide a focal plane with a diameter of 3.0~degrees, and a field-of-view (FOV) area of 7.06~square degrees. The Pan-STARRS2 (PS2) telescope hosts a 1.5~gigapixel camera and supports a slightly larger FOV. Each telescope is equipped with the same filter system, denoted as \grizy\ \citep[as described by][]{Tonry2012}. Images from both Pan-STARRS telescopes are processed with the Image Processing Pipeline \citep[IPP;][]{magnier2020a, waters2020}.

\begin{figure*}
    \centering
    \subfigure{\includegraphics[width=0.9\linewidth]{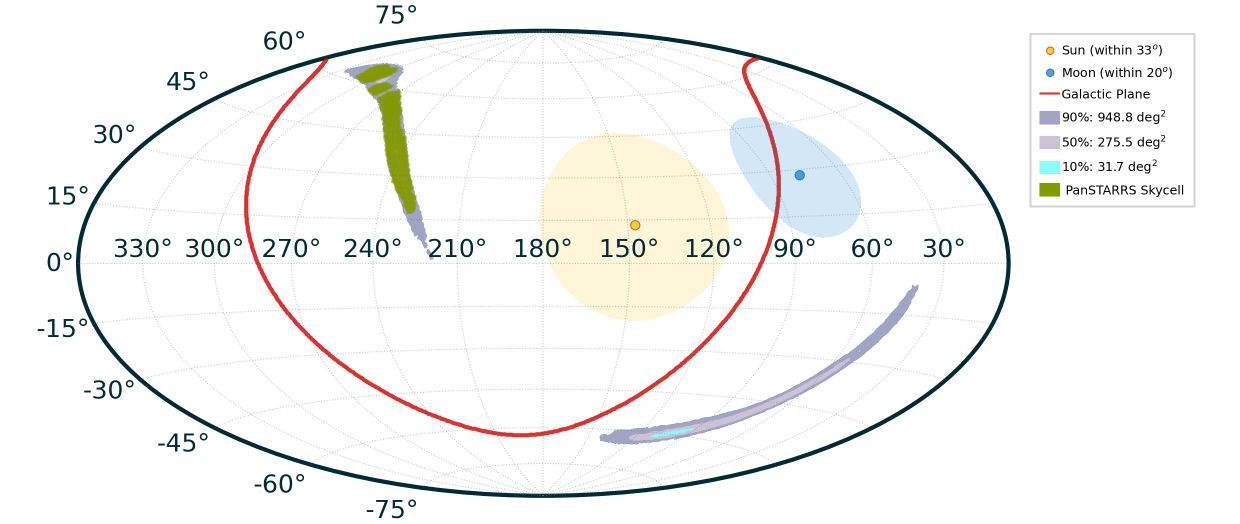}}
    \subfigure{\includegraphics[width=0.9\linewidth]{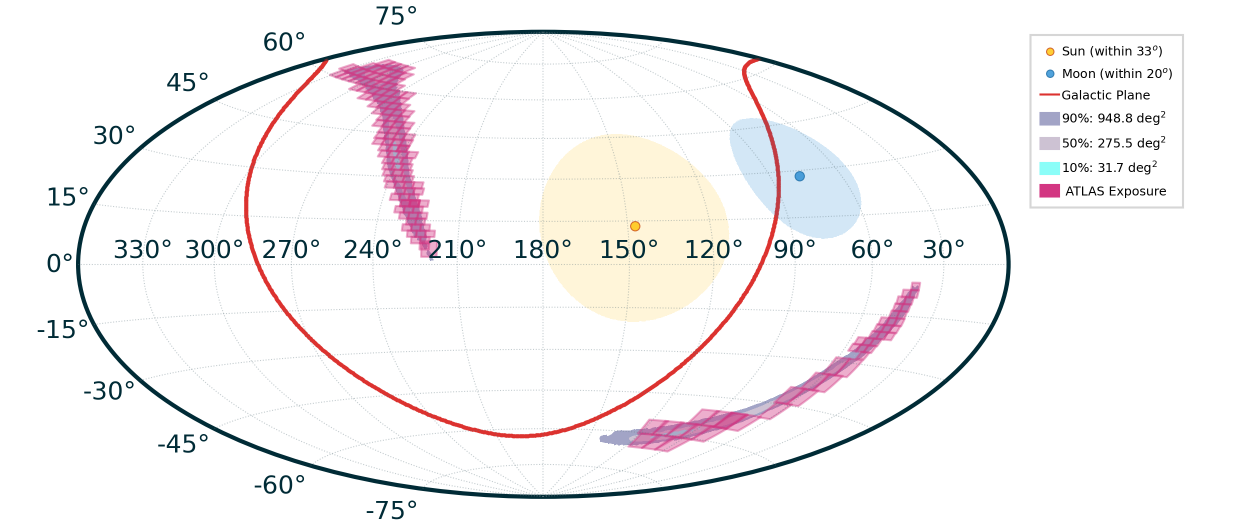}}
    \caption{
        \textit{Top:} The \whichMap\ skymap from \cite{LVKGCNDiscovery} with our targeted Pan-STARRS tiling observations performed within seven days post-burst overlaid.
        \textit{Bottom:} Same as above, but with the ATLAS observation footprints within seven days post-burst overlaid.
    }
    \label{FIG: Skymap}
\end{figure*}

The individual exposure frames (called warps) are astrometrically and photometrically calibrated \citep{magnier2020c} and overlapping exposures are co-added together with median clipping applied (to produce stacks). The Pan-STARRS1 $3 \pi$ reference sky images are subtracted from these frames \citep{waters2020} and photometry carried out on the resulting difference image \citep{magnier2020b}. These individual detections are ingested into the Pan-STARRS Transient Server database and assimilated into distinct objects with a time variable history. A series of quality filters are applied using the IPP image attributes and known asteroids and variable stars are removed. The objects remaining are cross-matched with all catalogued galaxies, AGN, CVs and historical transients \citep{Smith2020_ATLAS}, and simultaneously a machine-learning algorithm is applied to image pixel stamps at each transient position \citep{Wright2015, Smith2020_ATLAS}. This reduces the bogus detections to a manageable number for human scanning. For details on filtering and our human scanning process, specifically in the context of GW follow-up and faint transient identification, the reader is referred to \cite{Smartt2024} and \cite{Fulton2025}.

We began observing the LVK \whichMap\ skymap \citep{LVKGCNDiscovery} beginning on MJD~\MJDPSo, \MJDdiffo~days post-burst (night of 20 August 2025; HST). Our observations tiled the northern banana on three nights -- August 20, 21 and 22 -- with a series of dithered 45 and 120\,s exposures taken, reaching total exposure times per skycell of $90 - 3975$\,s (median exposure time of 360\,s) in the \ips-band. These images were combined into a single stacked image for each Pan-STARRS skycell \citep{Chambers2016arXiv_PanSTARRS1} and the standard processing as described above was immediately carried out. Typical $3.5 \sigma$ limiting magnitudes of $m_{i_{\textsc{ps}}} \simeq 19 - 23$ were achieved on these nightly stacks across the three nights, with the large range being due to varying sky conditions. We covered 286 and 318 square degrees (32\% and 34\%) of the \whichMap\ 90\% sky localisation region with Pan-STARRS within 3 and 7~days post-burst, respectively. We summarise our discoveries in Section~\ref{SEC: Candidate counterparts}. Our Pan-STARRS tiling of the \GWsource\ skymap is visually illustrated in Figure~\ref{FIG: Skymap}.

\subsection{ATLAS} \label{SEC: Scanning the skymap of S250818k - ATLAS}

The ATLAS system consists of five units, situated at Haleakala and Mauna Loa (Hawaii), El Sauce (Chile), Sutherland (South Africa), and Tenerife (Canary Islands) \citep{Tonry2018_ATLAS, Licandro2023_ATLASTeide}. Together, they tile the entire visible sky every $\sim 24$\,hrs (weather-dependent). The telescopes survey the sky in cyan ($c$) and orange ($o$), and the Tenerife unit uses a `wide' passband ($w$; CMOS response together with a UV/IR-cut filter, a similar effective wavelength to the combined Sloan $g'+r'$ bands). Typically, a pattern of $4 \times 30$\,s dithered exposures are acquired each night (roughly spaced logarithmically within a 1\,hr period). Our data processing pipeline follows standard practices; the images are photometrically and astrometrically calibrated using RefCat2 \citep{RefCat2}, before template subtraction is performed. As described by \cite{Smith2020_ATLAS}, quality cuts are applied, catalogued objects are cross-matched, and machine-learning algorithms are applied to the data to reject spurious detections and identify extragalactic transients \citep{2024RASTI...3..385W, 2025ApJ...990..201S}.

Similar to Pan-STARRS, we have a targeted GW follow-up program in place to rapidly tile the GW skymaps of candidate optical counterparts to GW signals \citep[e.g.,][]{Smartt2024}. Given the sub-threshold nature of the reported GW signal \GWsource, we did not initiate targeted follow-up searches. However, the ATLAS survey, during its routine sky survey mode, canvassed a substantial fraction of the \whichMap\ skymap within the first few nights post-burst. We note that during this period, ATLAS was undergoing a trial of extended exposures; as such, each ATLAS observation was composed of $4 \times 110$\,s exposures, resulting in deeper stacked target images compared with the `default' observing strategy (440 vs.\ 120\,s).

The ATLAS coverage within seven days post-burst is illustrated in Figure~\ref{FIG: Skymap}. We sampled 219, 604 and 831 square degrees (25\%, 65\% and 81\%) of the \whichMap\ 90\% sky localisation region with ATLAS within 1, 3 and 7~days post-burst, respectively. Each ATLAS pointing reached typical $5 \sigma$ limiting magnitudes of $m_o \simeq 18.8 - 20.4$ (110\,s) and $m_w \simeq 18.0 - 20.6$ (30\,s). Searching these data did not return any convincing candidate counterparts to \GWsource\ within the first three days after the GW signal.

\section{Candidate counterparts} \label{SEC: Candidate counterparts}

From our Pan-STARRS scanning of the northern portion of the \whichMap\ skymap on August 20, 21 and 22, we uncover 75 real extragalactic transient sources that lie within the 90\% contour of the northern `banana'. We found a fairly similar number of sources in our difference images that were faint and spatially coincident with the nucleus of a host galaxy, or coincident with a non-extended source that was either stellar or a compact galaxy. We visually inspected all and found no evidence of significant lightcurve changes (defined by $\Delta M \geq 0.2$\,mag\,d$^{-1}$ at more than $3 \sigma$ significance) that would imply a measurement of evolving luminosity. We deemed these to be likely due to AGN variability or subtraction artefacts, and thus did not consider them further. Our rejection of these is supported by the fact that $\lesssim 10$\% of short GRBs\footnote{Short GRBs are produced by compact binary mergers \citep{Kouveliotou1993, Berger2013, Tanvir2013}, and are thus a useful proxy for any candidate BNS merger counterpart.} occur within a host angular offset of 1'' \citep{O'Connor2022}. Of the 75 real extragalactic transient sources that are not obviously due to AGN variability, 28 had recorded activity prior to the GW burst, and thus are unrelated to \GWsource. The remaining 47 transients were reported to the IAU Transient Name Server (TNS),\footnote{\url{www.wis-tns.org}} and for these, we perform some additional investigation to determine their nature, and the likelihood of them being the optical counterpart associated with \GWsource. We searched for transients spatially associated with a host galaxy with a photometric\footnote{Photometric redshift estimates can be uncertain at these low redshifts, and so we only utilise these for the cases where spectroscopic redshift information is not available.} or spectroscopic redshift within the $2 \sigma$ limit from the LVK \whichMap\ skymap ($0.025 < z < 0.08$) that had no history of flux before \GWsource\ in our Pan-STARRS and ATLAS survey data, or in the ZTF public stream \citep[through the Lasair broker;][]{Lasair2024}, or registered on the TNS, and absolute magnitudes that place them in the KN regime \citep[$M \gtrsim -17$;][]{Metzger2010, Bulla2019, Nicholl2021}. These criteria reduced the list of viable candidates to 7 (excluding \SNName; for information on \SNName, see Section~\ref{SEC: SN2025ulz}). The full list of new transient sources discovered is listed in Table~\ref{TAB: Candidate list}, and we discuss our 7 viable candidates below. 

We draw attention to the limitations in our selection of the 7 candidates. It is possible that KNe (i)~may be brighter than $M \sim -17$ \citep[particularly if there is a long-lived neutron star;][]{YuZhangGao2013, MetzgerPiro2014}, (ii)~are nuclear, and/or (iii)~lie in galaxies that have a redshift within the $3 \sigma$ distance limit (rather than our $2 \sigma$ cut). While we considered these unlikely, we still list the sources we have discounted as counterparts in Table~\ref{TAB: Candidate list}. This table does not include the $\sim 70$ low-significance sources that are spatially coincident with the centres of their hosts and show no obvious signature of being a source of interest. If a GW counterpart lies within this sample we would be confusion limited, and further data (\eg, high-energy/radio emission, or a strong infrared excess) would be required to distinguish the counterpart from the background of confounding sources.

\subsection{Most viable candidates} \label{SEC: Candidate counterparts - Most viable candidates}

\textit{\ATxx{2025uso}/PS25grz}:
We first detected \ATxx{2025uso} on MJD 60907.3, the first night of our targeted follow-up campaign, at $m_i = 20.47 \pm 0.18$ \citep{Smartt2025GCN.41493}, coincident with a host galaxy with a spectroscopic redshift within the LVK range ($z_{\rm spec} = 0.025$; \citealt{DESI_DR1_DataRelease}). Although a promising candidate ($M_{i}\simeq-15$), further follow-up across the next two nights showed little lightcurve evolution. Furthermore, ATLAS detected \ATxx{2025uso} on MJD~60909.9 $\sim$~one magnitude brighter in $w$-band, and tracked a rise to peak at $m_o \sim 18$ over the following 20~days. Finally, the updated \whichMap\ skymap resulted in \ATxx{2025uso} lying outside the 90\% localisation region. Thus we discount \ATxx{2025uso} being related to \GWsource.

\textit{\ATxx{2025usy}/PS25gsa}:
We detected \ATxx{2025usy} on the first night of targeted follow-up observations (20 August 2025; MJD 60907.4) as a new transient with $m_i = 20.95 \pm 0.27$ \citep{Smartt2025GCN.41493}, coincident with a galaxy with a recorded photometric redshift, $z_{\rm phot} = 0.090 \pm 0.012$ \citep{LegacySurveys_DR9_DataRelease}, which would place it at $M_{i} \simeq -17$. The combination of the redshift of the host galaxy lying beyond the $2 \sigma$ LVK limit, the bright absolute magnitude, and the lightcurve remaining flat (\ie, within the errors, there is no evidence for a change in magnitude) across the subsequent two nights of observation (out to MJD 60909.4), led to us disfavouring \ATxx{2025usy} as a plausible candidate.

\textit{\ATxx{2025utr}/PS25gsh}:
\ATxx{2025utr} was discovered on MJD 60907.4 with $m_i = 19.84 \pm 0.13$ \citep{Smartt2025GCN.41493}, coincident with a galaxy with $z_{\rm phot} = 0.043 \pm 0.011$ \citep{LegacySurveys_DR9_DataRelease}. While a promising candidate (albeit on the brighter end) further observations up to MJD 60909.4 revealed a flat (within the errors) lightcurve evolution. We scheduled further observations with Pan-STARRS (300\,s $i$-band observation on MJD 60924.3) which confirmed the flat (or slightly rising) lightcurve evolution inferred from the earlier observations ($m_i = 19.67 \pm 0.03$). Finally, ATLAS detected \ATxx{2025utr} on MJD 60906.3, and observed its lightcurve rise in $w$ and $o$ across the next 15 days. The lightcurve resembles that of a supernova (\ie, its evolution follows the typical supernova $\sim 2 - 3$~week rise to peak), and thus we rule out \ATxx{2025utr} as a kilonova candidate counterpart to \GWsource.

\textit{\ATxx{2025utx}/PS25gsm}:
\ATxx{2025utx} was discovered on MJD 60907.4 at $m_i = 20.21 \pm 0.26$. It is coincident with a galaxy with $z_{\rm spec} = 0.080$ \citep{DESI_DR1_DataRelease}, which places it right at the upper $2 \sigma$ bound for the LVK \whichMap\ skymap localisation \citep{LVKGCNDiscovery}. At this redshift, \ATxx{2025utx} possessed an absolute mag, $M_i = -17.6$, which is likely too bright for a KN, and further observations with Pan-STARRS (300\,s $i$-band observation on MJD 60924.3) confirmed the flat lightcurve evolution inferred from the earlier observations ($m_i = 20.12 \pm 0.11$). The flat lightcurve and bright magnitude compared to what we might expect for kilonovae rules out \ATxx{2025utx} as a suitable candidate counterpart.

\textit{\ATxx{2025uuf}/PS25gsv}:
\ATxx{2025uuf} was detected on MJD 60907.4 at $m_i = 20.99 \pm 0.29$ \citep{Smartt2025GCN.41493}. It is coincident with a galaxy with $z_{\rm phot} = 0.061$ \citep{Bilicki2014_2MASS}, within the LVK distance range. The inferred absolute magnitude of $M_i \sim -16$ would put it in the kilonova regime, but subsequent observations across the next two nights (up to MJD~60909.6) revealed no sign of decline, and a flat lightcurve evolution. Further observations with Pan-STARRS (300\,s $i$-band observation on MJD~60924.3) confirmed the flat (or slightly rising) lightcurve evolution inferred from the earlier observations ($m_i = 20.58 \pm 0.11$), and thus we suggest  \ATxx{2025uuf} is likely a supernova, and not associated with \GWsource.

\textit{\ATxx{2025uuw}/PS25gtm}:
We detected \ATxx{2025uuw} on MJD 60907.4 with an initial AB magnitude, $m_i = 20.44 \pm 0.20$ \citep{Smartt2025GCN.41493}, coincident with a galaxy with a photometric redshift, $z_{\rm phot} = 0.034 \pm 0.006$ \citep{LegacySurveys_DR9_DataRelease}. Follow-up observations on MJD~60909.3 revealed a rising source and further observations with Pan-STARRS (300\,s $i$-band observation on MJD 60924.2) showed substantial brightening from the previous epochs of observation, with the $i$-band magnitude rising by more than two magnitudes ($m_i = 18.10 \pm 0.02$). ATLAS detected \ATxx{2025uuw} on MJD 60911.0 in the $w$-band, and tracked its brightening in $w$ and $o$. Finally, \ATxx{2025uuw} was classified as SN~Ia at $z = 0.03$ \citep{2025TNSCR3487....1F}. 

\textit{\ATxx{2025uxs}/PS25guo}:
We discovered \ATxx{2025uxs} on MJD 60909.3, the third night of our targeted follow-up observations \citep{Smartt2025GCN.41493}. It was detected with $m_i = 20.05 \pm 0.08$, and associated with a galaxy with $z_{\rm spec} = 0.063$ \citep{DESI_DR1_DataRelease}, giving \ATxx{2025uxs} an absolute AB mag, $M_i = -17.2$. Further observations with Pan-STARRS (300\,s $i$-band observation on MJD 60924.2) showed substantial brightening ($m_i = 18.48 \pm 0.02$) and ATLAS data showed a rising lightcurve in $w$ and $o$ for the subsequent 15~days. The lightcurve again resembles a supernova, and thus we conclude that \ATxx{2025uxs} is not a kilonova-like source associated with \GWsource.

In summary, from our targeted skymap tiling with Pan-STARRS, and routine ATLAS observations, we uncover no convincing optical kilonova-like counterpart candidates to the GW event \GWsource. However, we note that we were only able to cover the northern banana with Pan-STARRS, and our follow-up campaign commenced $\approx 2.2$~days post-burst, and so we may not have been sensitive to an optical counterpart, were it faint and rapidly evolving (a point we consider further in Section~\ref{SEC: KN limits from Pan-STARRS}).

\subsection{Alternative counterpart scenario} \label{SEC: Candidate counterparts - Alternative counterpart scenario}

\cite{Metzger2024} and \cite{Chen2025} present a scenario to explain the potential of sub-Solar-mass NS (ssNS) mergers, possibly detectable via gravitational waves by LVK. In this scenario, ssNSs are formed in the gaseous accretion disk surrounding a young collapsar event. These ssNSs then potentially undergo multiple mergers on a rapid timescale (minutes -- hours; at most $\leq$~days), with each merger possibly producing a detectable GW signal. Interestingly, the optical counterparts to these systems will not appear as a `typical' kilonova signal; any signal powered from \rpro\ radioactive decay will be dwarfed by the radiation emitted from the collapsar. Thus, it may be the case that the optical signal one should search for as the counterpart to ssNS merger GW signals is that of a collapsar (possibly with some boosted emission from the kilonova(e) within the system).

We performed a search for a coincident collapsar signal in our ATLAS and Pan-STARRS (and all publicly available) data. Specifically, we scanned all TNS-registered events for SN-like transients that were located within the \whichMap\ skymap (including those discussed above), associated with a galaxy within the $2 \sigma$ distance limit of the GW signal, and with a plausible explosion epoch $\lesssim 3$~days post-GW signal. From this search, we do not uncover any convincing counterpart signal. This is perhaps not so surprising, given that one of the predictions of the scenario presented by \cite{Metzger2024} and \cite{Chen2025} is that a sequence of merger events may occur in quick succession, producing multiple, independent, GW signals. No subsequent related GW events were reported prior to, or following, \GWsource.

We note that future optical follow-up campaigns to GW triggers (especially those for ssNS mergers) should consider this merger scenario, as it provides an alternative, distinct optical signature for which one can search. In this scenario, we would not be searching for rapidly fading, faint optical transients (on timescales of $\sim$~days), but rather rising SN-like sources that would need to be followed and quantified for many weeks.

\section{\SNName} \label{SEC: SN2025ulz}

\subsection{Initial discovery and observations} \label{SEC: SN2025ulz - Initial discovery and observations}

AT\,2025ulz was  discovered by the Zwicky Transient Facility \citep[ZTF;][]{Bellm2019} during its follow-up observations tiling the localisation region of \GWsource, with AB magnitudes, $m_g = 20.99 \pm 0.13$ and $m_r = 21.29 \pm 0.13$ \citep{Stein2025GCN.41414}.  The ZTF coverage of the skymap began quickly, just 2.7\,hrs after the GW trigger \citep{Stein2025GCN.41414}, and follow-up observations revealed that AT\,2025ulz was fading quickly and changing colour \citep{Busmann2025GCN.41421, Hall2025GCN.41433}. While the position of AT\,2025ulz was covered in our Pan-STARRS tiling of the map of \GWsource, we also triggered targeted follow-up observations of AT\,2025ulz with Pan-STARRS in \grizy\ with longer exposures, due to the report of $m_r = 21.8$\,mag from \cite{Hall2025GCN.41433} and its fading nature. 

We commenced observing on the night of 20 August 2025 (MJD~60907.3), approximately 2.2~days post-burst \citep{Gillanders2025GCN.41454}. We continued nightly multi-colour monitoring of AT\,2025ulz alongside our follow-up program tiling the \GWsource\ skymap (as outlined in Section~\ref{SEC: Scanning the skymap of S250818k - Pan-STARRS}). We began with a nightly cadence until $\simeq 8$\,d post-burst, where we then relaxed to a $2 - 4$\,d cadence. We dropped the $zy$-bands after observations on MJD~60908.2 returned only upper limits. \cite{ENGRAVE2025GCN.41476} obtained a spectrum of AT\,2025ulz on MJD~60911.0 and classified the transient as a SN IIb. Despite this classification, we continued our monitoring of \SNName\ (now classified as a SN on the TNS) as it presented a good opportunity to sample and characterise a source that, at least initially, seemed a promising candidate counterpart to \GWsource. 

We also undertook targeted observations of \SNName\ with the 40\,cm SLT and Lulin One-meter Telescope (LOT) at the Lulin observatory, in SDSS $ugri$ filters as part of the Kinder collaboration \citep[][]{Chen2025_SN2024ggi}. These images were calibrated using a custom-built pipeline \citep{LSSTpipeline}, and photometry was measured using the \textsc{AutoPhOT} pipeline \citep{Brennan2022_AUTOPHOT}. A full summary of our follow-up photometric observations is presented in Table~\ref{TAB: SN2025ulz photometry} and illustrated in Figure~\ref{FIG: SN2025ulz lightcurve}.

At the Pan-STARRS sky location of \SNName\ (\PSCoords), the STScI PS1 $3 \pi$ stamp server reference frames are shallower than our stacked target images \citep{Flewelling2020}. Subtraction of a reference template shallower than the science image can lead to substantial systematic errors in measuring reliable photometry for the transient. Specifically, we find that use of the PS1 $3 \pi$ $i$-band reference images result in systematic offsets from the true photometry of order $\lesssim 0.8$\,mag. Since the release of the PS1 DR2 images at STScI, both PS1 and PS2 have continued observing, and in certain regions of the sky we can construct significantly deeper reference images from this proprietary data. Our new stacked reference frames sum to total exposure times for each of \izy\ of $t_i = 4445$, $t_z = 6720$ and $t_y = 2760$~seconds.\footnote{We do not possess many historical $gr$-band images, and so rely on the original DR2 templates, with total exposure times of $t_g = 817$ and $t_r = 960$~seconds.} Using these as reference templates allows us to extract accurate photometry for \SNName, as well as ensuring reliable host galaxy subtraction, a vital step given the proximity of the transient to the host galaxy nucleus. To enable robust analysis of \SNName\ by other groups, we provide public access to these improved reference frames (see Data Availability).

\subsection{Historical activity search in Pan-STARRS and ATLAS} \label{SEC: SN2025ulz - Historical activity search}

To check for any prior transient or variable activity of this source, we searched the location of \SNName\ in our Pan-STARRS data. We uncover two recent non-detections ($m_i > 21.0$ at 9.7\,d pre-burst, and $m_y > 19.7$ at 0.82\,d pre-burst; both $3 \sigma$ upper limits) at the location of \SNName\ (first reported by \citealt{Nicholl2025GCN.41439}). These data, in particular the $y$-band upper limit, place useful constraints on the presence of a bright transient at the location of \SNName\ immediately prior to the GW merger.

\begin{figure*}
    \centering
    \includegraphics[width=0.8\linewidth]{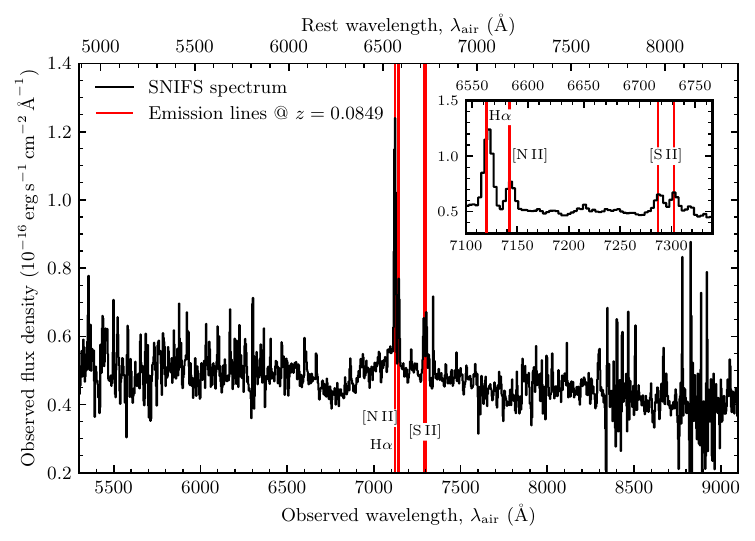}
    \caption{
        SNIFS spectrum of the host galaxy of \SNName. We mark the location of prominent
        H$\alpha$, [N\,\II]~$\lambda6583.46$, [S\,\II]~$\lambda \lambda 6716.44 \ \& \ 6730.81$ emission, which allows us to measure a host galaxy redshift, $z = 0.0849 \pm 0.0003$.
    }
    \label{FIG: Host galaxy spectrum}
\end{figure*}

Despite not being as sensitive as the Pan-STARRS survey, ATLAS can still provide useful constraints on pre-merger activity, especially given its high-cadence sky coverage. We forced flux measurements at the sky position of \SNName\ in all ATLAS difference images for a period of \ATLASforceddays\ days before the discovery of \GWsource\ (using the ATLAS Forced Photometry Server; \citealt{Shingles2021}). This includes 28 separate nights of observations by the ATLAS telescopes during that time. The individual 30 and 110\,s (see Section~\ref{SEC: Scanning the skymap of S250818k - ATLAS}) frame fluxes and errors were combined into one nightly measurement. We can confirm that no previous source existed at this position within the ATLAS data down to a $3 \sigma$ limiting magnitude, $m_o \sim 19.8 - 21.6$. The most recent pre-GW trigger ATLAS observation of the location of \SNName\ was performed at MJD 60903.322 (1.73\,d pre-burst), and from these pointings, we were able to obtain a $3 \sigma$ limiting magnitude of $m_o > 21.6$ \citep{Srivastav2025GCN.41451}. ATLAS also observed the field on MJD 60905.916 (0.86\,d post-burst), yielding a $3 \sigma$ upper limit of $m_w > 21.7$ at the position of \SNName.

\subsection{Redshift estimation} \label{SEC: SN2025ulz - Redshift estimation}

\SNName\ is offset from the centre of its host galaxy \HostGalName\ by 0.88''. At the time of discovery of \SNName\ by ZTF, only a photometric redshift, $z_{\rm phot} = 0.091 \pm 0.016$ \citep{LegacySurveys_DR9_DataRelease}, was available. We performed an observation of the host galaxy with the SNIFS instrument (SuperNova Integral Field Spectrograph; \citealt{Lantz2004_SNIFS}) on the University of Hawaii 2.2\,m telescope at Maunakea, beginning on 2 September 2025 with an exposure time of 2700~seconds. SNIFS has two channels, split by a dichroic mirror, spanning $3400 - 5100$\,\AA\ and $5100 - 10000$\,\AA, with average spectral resolutions of 5 and 7\,\AA\ for the Blue and Red channels, respectively. The data was processed with the quick nightly reduction pipeline described by \cite{Tucker2022}.

In this spectrum (Figure~\ref{FIG: Host galaxy spectrum}) we identify prominent emission features centred at 7119, 7142, 7286 and 7302\,\AA, which correspond to galaxy emission lines of H$\alpha$, [N\,\II] $\lambda6583.46$, [S\,\II] $\lambda \lambda 6716.44 \ \& \ 6730.81$, respectively, at a redshift, $z = 0.0849 \pm 0.0003$. This measurement for the host redshift is in agreement with other measurements reported by \cite{Karambelkar2025GCN.41436} and \cite{ENGRAVE2025GCN.41476}.

For our chosen cosmological parameters (see Section~\ref{SEC: Introduction}), the redshift of \SNName\ corresponds to a luminosity distance, $D_{\rm L}$~\HostDist. This lies just within the reported $2 \sigma$ limit for the distance in the \whichMap\ skymap in the direction of \SNName, which is $D_{\rm L} = 266 \pm 70$\,Mpc.

\begin{figure*}
    \centering
    \subfigure{\includegraphics[width=\linewidth]{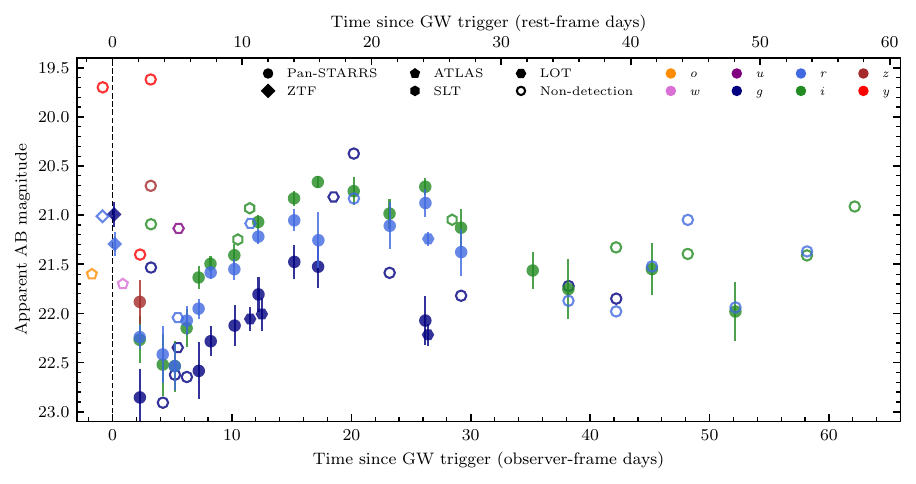}}
    \caption{
        Lightcurve of \SNName. We include the discovery $gr$-band data from ZTF \citep{Stein2025GCN.41414} in addition to our multi-colour Pan-STARRS, ATLAS, SLT and LOT data. Hollow symbols correspond to $3 \sigma$ upper limits.
    }
    \label{FIG: SN2025ulz lightcurve}
\end{figure*}

\subsection{Observed properties of \SNName} \label{SEC: SN2025ulz - Observed properties}

The first peak of \SNName\ is short-lived, with its maximum occurring before our first Pan-STARRS observation. The discovery $gr$-band magnitudes from ZTF \citep{Stein2025GCN.41414}, acquired just 0.13 and 0.19\,d after the GW trigger, at $m_g = 20.99 \pm 0.13$ and $m_r = 21.29 \pm 0.13$, respectively \citep{2025TNSTR3264....1S}, are the brightest report in GCNs and on the TNS from this early phase, and so we assume them to be at (or close to) peak. We measured \SNName\ to rapidly fade from this peak to $m_g = 22.85 \pm 0.29$ and $m_r = 22.24 \pm 0.12$ in our first Pan-STARRS $gr$-band observations acquired at 2.29 and 2.27\,d post-merger, respectively, corresponding to decline rates of 0.86\,mag\,d$^{-1}$ ($g$) and 0.45\,mag\,d$^{-1}$ ($r$). For comparison, \gfo\ had a $g$-band ($r$-band) fade rate of 1.55\,mag\,d$^{-1}$ (1.15\,mag\,d$^{-1}$) from 1.4 to 2.4~days post-explosion (see Section~\ref{SEC: SN2025ulz - Comparison to AT2017gfo} and Figure~\ref{FIG: SN2025ulz + AT2017gfo + KN models}). The colour of \SNName\ evolved rapid in this early phase too, from $g - r \approx -0.3$ at discovery, to $g - r \approx +0.6$ in the Pan-STARRS observations at 2.3\,d.

The Milky Way extinction toward the line of sight of \SNName\ is $E (B - V) =$~\EBsubV, and in the Pan-STARRS1 filters (assuming $R_{V} = 3.1$) this corresponds to $A_g =$~\Ag, $A_r =$~\Ar, $A_i =$~\Ai, $A_z =$~\Az, $A_y =$~\Ay\ and $A_w =$~\Aw~AB mag \citep{Schlafly2011}, which we use for calculation of absolute magnitudes.

\begin{figure*}
    \centering
    \subfigure{\includegraphics[width=\linewidth]{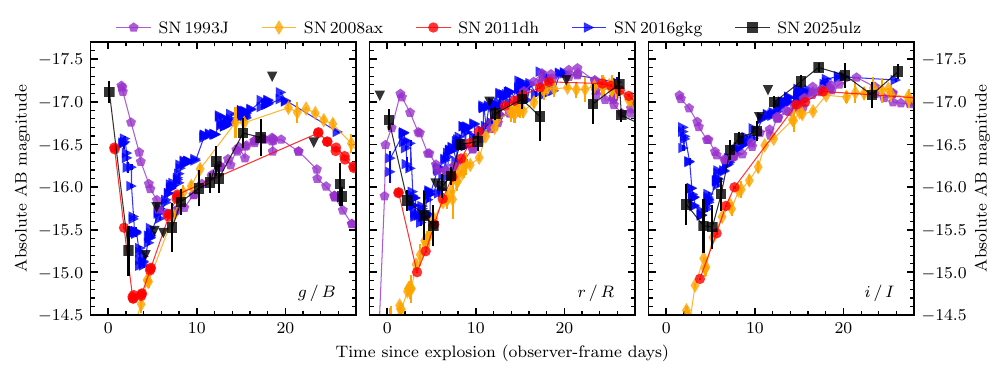}}
    \subfigure{\includegraphics[width=\linewidth]{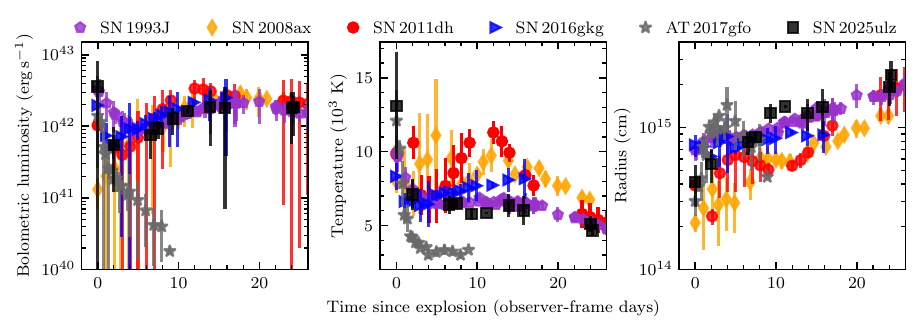}}
    \caption{
        \textit{Top:} Comparison of \SNName\ $gri$-band evolution to the early lightcurves of SNe 1993J ($BRI$), 2008ax ($gri$), 2011dh ($gri$) and 2016gkg ($gri$), all SNe IIb that possess early-time observations. The evolution of \SNName\ closely resembles the evolution of these transient events.
        \textit{Bottom:} Bolometric luminosity ($L_{\rm BOL}$) lightcurve property comparisons between \SNName, SNe 1993J, 2008ax, 2011dh and 2016gkg, and the KN \gfo. The $L_{\rm BOL}$ evolution of \SNName\ closely mirrors the comparison sample of SNe IIb, whereas it exhibits a stark difference to that of \gfo. A similar trend exists for the temperature comparison.
    }
    \label{FIG: Comparisons to SNe IIb + bolometric luminosity comparisons}
\end{figure*}

Adopting only Milky Way extinction, the absolute magnitude of the source faded from  $M_g \simeq -17.1$ and $M_r \simeq -16.8$ at discovery to our Pan-STARRS observations at 2.3\,d of $M_g \simeq -15.3$, $M_r \simeq -15.8$, $M_i \simeq -15.8$ and $M_z \simeq -16.2$. The full optical lightcurves, in absolute magnitudes, are presented in Figures~\ref{FIG: Comparisons to SNe IIb + bolometric luminosity comparisons}~and~\ref{FIG: SN2025ulz + AT2017gfo + KN models}.

From just the first few epochs of data, the evolution of \SNName\ qualitatively resembles the evolution of the kilonova \gfo\ (see Figure~\ref{FIG: SN2025ulz + AT2017gfo + KN models} and Section~\ref{SEC: SN2025ulz - Comparison to AT2017gfo}). It is around 1\,mag brighter at peak, although it was discovered earlier compared to the GW trigger than \gfo.  With data from just the first 2.5~days in hand, temporal  coincidence with \GWsource, and 3D spatial consistency, an optical counterpart claim was reasonable \citep{Karambelkar2025GCN.41436}. However, continued observations with Pan-STARRS reveal an upturn in the lightcurve after $\sim 5$~days, which is unexpected for any observed or modelled kilonova. The slow, continuous rise for the subsequent $\approx 15$~days resemble a SN lightcurve.

\subsection{Comparison to SNe IIb} \label{SEC: SN2025ulz - Comparison to SNe IIb}

The lightcurve of \SNName\ begins to re-brighten at a time $t \gtrsim 5$\,d from the discovery epoch (see Figure~\ref{FIG: SN2025ulz lightcurve}), and continues for a further 10~days. The early, luminous peak, followed by rapid fading, and subsequent second rise to a peak $\sim 15$\,d post-explosion is reminiscent of SN~IIb lightcurve behaviour. To further investigate this, we compare the multi-band lightcurve of \SNName\ with well-sampled lightcurves of the SNe~1993J \citep[$BRI$;][]{1994AJ....107.1022R}, 2008ax \citep[$gri$;][]{2008MNRAS.389..955P}, 2011dh \citep[$gri$;][]{Arcavi2011} and 2016gkg \citep[$gri$;][]{2017ApJ...837L...2A, Tartaglia2017}. We correct all observed photometry for reddening effects before converting to absolute magnitudes, for ease of comparison. For SNe~1993J, 2008ax, 2011dh and 2016gkg, we consult the above references for extinction corrections, distance and explosion time estimates. For \SNName, we assume an explosion phase equal to the GW merger time \citep{LVKGCNDiscovery}. This is to show that even if the GW merger time was close to the SN explosion by coincidence, the lightcurve still resembles a typical SN~IIb. 

In all $gri$-bands, the qualitative evolution of \SNName\  mirrors SNe~2011dh and 2016gkg. The data resemble the evolution of \SNxx{1993J}, however it appears that \SNName\ evolved on a more rapid timescale. It is also very similar to the high-cadence lightcurves of \SNxx{2017jgh} and \SNxx{2021zby}, SNe~IIb observed by Kepler and TESS, respectively \citep{Armstrong2021_17jgh, Wang2023_21zby}. However, a direct comparison of \SNName\ to these two SNe is more challenging, since the Kepler and TESS filters are extremely broad. These shock-cooling tails are common in SNe~IIb, but vary in their duration and peak luminosity \citep{2025A&A...701A.128A}, and are modelled with extended envelopes of modest mass \citep[e.g.,][]{2017ApJ...837L...2A, 2017ApJ...846...94P}. In some cases, such as \SNxx{2008ax}, no shock-cooling signature is observed. The absolute magnitudes of all comparison SNe agree well with \SNName; both in the $^{56}$Ni-powered main peak, as well as the initial sharp decline. The evolutionary timescales (\SNxx{1993J} aside) also agree well. From this visual inspection, it is clear that, based on photometry alone, the photometric data can be well-explained by \SNName\ being a typical SN IIb. Furthermore, despite only correcting for Milky Way extinction, the data of \SNName\ closely match these comparison events. This indicates that host galaxy extinction for \SNName\ may be minor.

To further quantify the agreement between \SNName\ and our sample of comparison SNe~IIb, we fit the photometry of each event with \texttt{SuperBol} \citep{SUPERBOL} to extract bolometric luminosity ($L_{\rm BOL}$) lightcurve properties. In Figure~\ref{FIG: Comparisons to SNe IIb + bolometric luminosity comparisons}, we compare the bolometric lightcurves and photospheric temperatures and radii extracted for \SNName\ with those of SNe 1993J, 2008ax, 2011dh and 2016gkg, as well as the KN \gfo. We find that the $L_{\rm BOL}$ evolution of \SNName\ closely matches all comparison SNe. The evolution of \gfo\ is starkly different, plummeting in luminosity much faster than any of the SNe. A similar trend is found for temperature evolution. While there is some scatter among the population of SNe~IIb (and \SNName), the evolution of \gfo\ is again distinct, as it exhibits much more rapid cooling than any of these other events. These $L_{\rm BOL}$ comparisons again unambiguously demonstrate the similarities between \SNName\ and SNe~IIb, while also highlighting the differences with the KN \gfo\ (see Section~\ref{SEC: SN2025ulz - Comparison to AT2017gfo} for more comparison to \gfo).

The SNIFS observation described in Section~\ref{SEC: SN2025ulz - Redshift estimation} was centred on the position of \SNName, and excess flux is visible at the transient position in the collapsed white light B and R cubes, offset from the galaxy core. The seeing was excellent ($0.6 - 0.8$~arcsec), and the SNIFS spaxals subtend 0.43~arcsec. We extracted a spectrum within a 0.8~spaxal radius of the \SNName\ position. The spectrum (taken 15~days post-GW signal; see Figure~\ref{FIG: Host galaxy spectrum}) possesses broad absorption to the blue of H$\alpha$, which likely corresponds to P-Cygni absorption from \SNName, as described in the spectroscopic classification report of \SNName\ as a SN of type II or IIb \citep{ENGRAVE2025GCN.41476}. The velocities are consistent between the VLT+MUSE spectrum \citep{ENGRAVETNSClass} and our SNIFS spectrum, indicating a velocity for the P-Cygni trough of $v_{\rm ej} \simeq 14000$\,km\,s$^{-1}$. This is fairly typical for a SN~IIb, with \SNxx{2011dh} showing H$\alpha$ absorption trough minima at velocities of $16000 - 13000$\,km\,s$^{-1}$ between explosion and 15 days \citep{2014A&A...562A..17E}. We suggest that the consistency between the ENGRAVE collaboration's spectral typing and our SNIFS spectrum rules out \SNName\ being a kilonova, and favours a supernova of type~IIb.

\subsection{Comparison to \gfo} \label{SEC: SN2025ulz - Comparison to AT2017gfo}

\begin{figure*}
    \centering
    \subfigure{\includegraphics[width=0.8\linewidth]{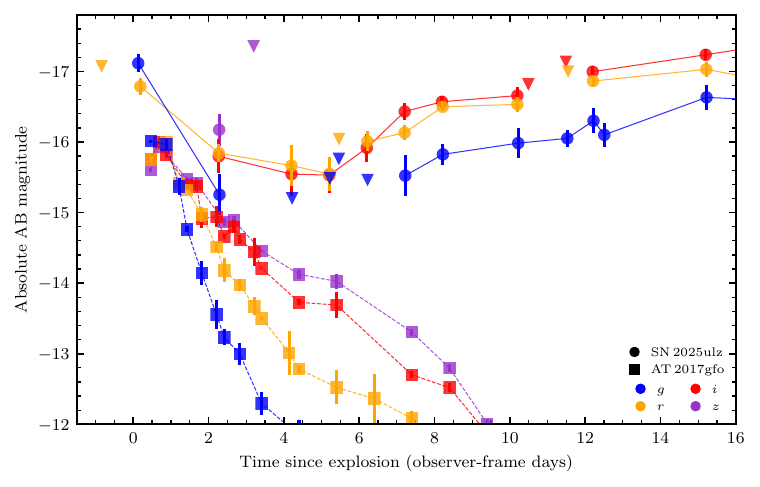}}
    \includegraphics[width=0.495\linewidth]{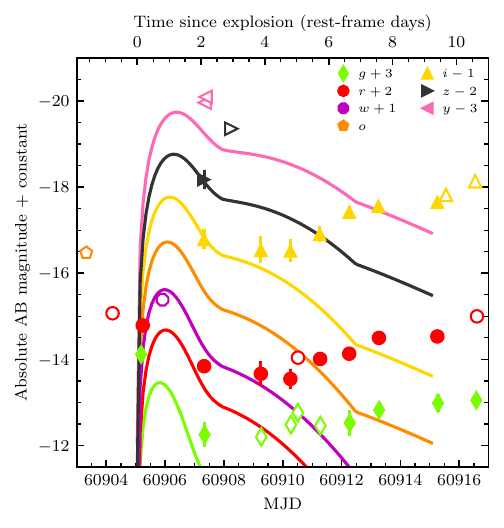}
    \includegraphics[width=0.495\linewidth]{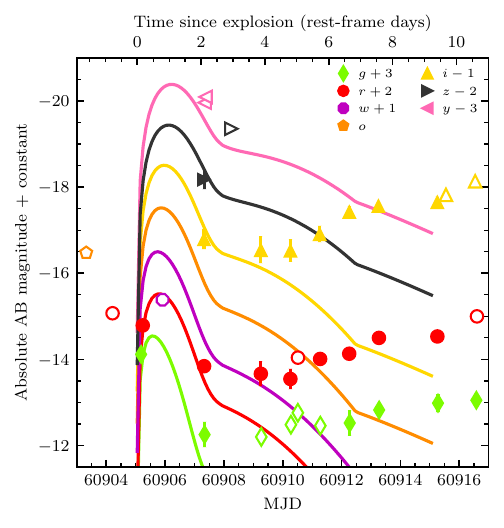}
    \caption{
        \textit{Top:} Comparison of the $griz$-band lightcurves of \SNName\ and \gfo. Downward-facing triangles correspond to $3 \sigma$ upper limits for \SNName.
        \textit{Bottom:} KN model fits for \SNName. Both models presented have dynamical (low opacity) ejecta with 0.025\,\msun\ and 0.3\,c, and disk wind (higher opacity) ejecta with 0.05\,\msun\ and 0.15\,c. The model on the left possesses no shock heating, whereas the model on the right is generated invoking maximum shock heating. Hollow symbols correspond to $3 \sigma$ upper limits.
    }
    \label{FIG: SN2025ulz + AT2017gfo + KN models}
\end{figure*}

To further demonstrate the incompatibility of a KN interpretation, we compare the optical $griz$-band lightcurve of \SNName\ with \gfo. Specifically, we compare to the optical data from \cite{Andreoni2017, Arcavi2017, Chornock2017, Cowperthwaite2017, Drout2017, Evans2017, Kasliwal2017, Pian2017, Smartt2017, Tanvir2017, Troja2017, Utsumi2017, Valenti2017}, compiled by \cite{Coughlin2018}.\footnote{This photometry data file is available at \url{www.engrave-eso.org}.} As above, we correct the data for reddening effects, before converting into absolute magnitudes \citep[assuming a distance of 40.4\,Mpc;][]{Hjorth2017}. This comparison is plotted in Figure~\ref{FIG: SN2025ulz + AT2017gfo + KN models}.

The rapid decline from the $gr$-band ZTF discovery points to the first $griz$-band data from Pan-STARRS at 2.3\,d post-GW trigger is not as fast as that observed for \gfo\ at the same phases (although it is somewhat comparable; see Section~\ref{SEC: SN2025ulz - Observed properties}). \SNName\ is $0.5 - 1$\,mag brighter than \gfo\ in both $g$ and $r$ at $\sim 0.5$\,d post-explosion. Beyond $\approx 2.3$~days, \gfo\ continues to evolve on a rapid timescale across all $griz$-bands, whereas \SNName\ flattens. The re-brightening in the $gri$-band data of \SNName\ is irrefutable after $\approx 6$~days, at which point \SNName\ is $\gtrsim 2.5$\,mag brighter in $i$-band (the discrepancy is even larger in $gr$-bands). Clearly, the evolution of \SNName\ does not match that of \gfo\ beyond a few days. This is also clear from our $L_{\rm BOL}$ lightcurve comparisons in Figure~\ref{FIG: Comparisons to SNe IIb + bolometric luminosity comparisons}. However, kilonovae are expected to exhibit a diverse range of observational properties, and as such, one cannot formally rule out the KN interpretation given the data do not exactly match \gfo. Thus, we undertake a modelling approach to see if the lightcurve of \SNName\ can be reproduced by a KN.

\subsection{Fiducial kilonova model} \label{SEC: SN2025ulz - Fiducial kilonova model}

As noted by \cite{Stein2025GCN.41414, Busmann2025GCN.41421, Hall2025GCN.41433} and shown in Figure~\ref{FIG: SN2025ulz lightcurve}, the early emission from \SNName\ indicated rapid fading, which could be consistent with expectations for a kilonova. We now consider whether any plausible kilonova model could fit the observed optical lightcurve. We explore whether a kilonova interpretation can be excluded based only on the first few detections (similar to the events discussed in Section~\ref{SEC: Candidate counterparts}), whether full lightcurve monitoring is necessary, or whether spectroscopy (unambiguously identifying \SNName\ as a supernova) is the only way to rule out a kilonova in this instance.

We attempted to model the lightcurve from Pan-STARRS, along with the ZTF discovery points, using the \texttt{Modular Open Source Fitter for Transients} \citep[\texttt{MOSFiT};][]{guillochon2018mosfit}. We use the analytic (Arnett-like; \citealt{1982ApJ...253..785A}) radioactively heated kilonova model from \citet{Metzger2017}, as implemented by \cite{Villar2017}, with additional luminosity from shock cooling (following \citealt{Piro2018} and \citealt{Nicholl2021}). This is the same model used by \citet{Rastinejad2022} to analyse the kilonova associated with \GRBxx{211211A}. Applying this to \SNName, we were unable to find a converged model that gives a reasonable fit to the full lightcurve, trivially confirming that the source is not a kilonova. Next, we build a fiducial model to match the early peak to demonstrate when the data and models diverge. 

We use a two-component ejecta: `blue', with opacity $\kappa_{\rm b} = 0.5$\,cm$^2$\,g$^{-1}$ and velocity 0.3\,c, and `purple', with opacity $\kappa_{\rm p} = 3$\,cm$^2$\,g$^{-1}$ and velocity 0.15\,c. These roughly correspond to the dynamical (blue) and disk wind (purple) ejecta found in \texttt{MOSFiT} fits to \gfo\ \citep{Villar2017, Nicholl2021}. These two components dominate the luminosity on timescales of $\sim 1$~day and $\sim 1$~week, respectively. To reproduce the fast fade between the first detection from ZTF and our earliest Pan-STARRS observation, a blue ejecta mass of $\approx 0.025$\,\msun\ is required. This is comparable to the mass inferred for \gfo. We also assume a purple ejecta mass of 0.05\,\msun, based on \gfo.

The results are shown in Figure~\ref{FIG: SN2025ulz + AT2017gfo + KN models}. With only \rpro\ heating, the model luminosity falls short of the initial ZTF detections by $1 - 2$\,mag, but provides a reasonable match to the Pan-STARRS data (particularly in the redder bands) two days later. With the addition of shock heating, we can also match the early blue emission. Doing so requires setting the parameter \texttt{shock\_frac}~$=1$; i.e., all the dynamical ejecta are heated by a strong shock, which could for example result from a GRB jet punching through kilonova ejecta \citep{Piro2018}. Analysis of \gfo\ with a binary-constrained kilonova model (i.e., ejecta mass is determined by the masses of the constituent NSs) also showed evidence for shock heating \citep{Nicholl2021}, though with a lower \texttt{shock\_frac}~$\approx 0.5$. The early UV emission from \GRBxx{211211A} suggested a \texttt{shock\_frac}~$= 0.6 - 1$, though that event had an unusually long-lived GRB jet. In summary, the optical lightcurve of \SNName\ during the first two days is consistent with a kilonova model that is perhaps somewhat extreme, but compatible with previous kilonova models.

However, the model is unable to account for the flattening, and subsequent re-brightening, of the lightcurve at later times. While the purple component can produce emission on longer timescales, visible as a bump in the model lightcurves at $\sim 3 - 5$~days, increasing the mass of this slower, redder component would also increase the luminosity on day two, and exceed the observed brightness in the redder bands ($i,z$, and especially $y$). Adding an extremely long-lived and redder component (which could arise from tidal ejecta in an asymmetric binary) to try to reproduce the slow rise would require an unrealistic ejecta mass (possibly orders of magnitude larger than in \gfo, and outside the range where the \texttt{MOSFiT} model is valid). We conclude that no reasonable kilonova model could reproduce the lightcurve evolution beyond $\sim 5$~days.

\subsection{Host galaxy SED and modelling} \label{SEC: SN2025ulz - Host galaxy SED and modelling}

\SNName\ is associated with the galaxy \HostGalName, with coordinates, RA~=~15:51:54.156, Dec~=~+30:54:09.24. To characterize the host galaxy, we make use of the \texttt{HostPhot} tool \citep{HostPhot}, which extracts consistent multi-band photometry from archival surveys. We obtained cutouts from Pan-STARRS, SDSS, Legacy Survey, GALEX, 2MASS and unWISE, covering rest-frame wavelengths from the far-ultraviolet through the mid-infrared. A minimum error floor of 0.05\,mag was added in quadrature to account for calibration systematics, and galactic foreground extinction was corrected.

\begin{figure*}
    \centering
    \includegraphics[width=\textwidth]{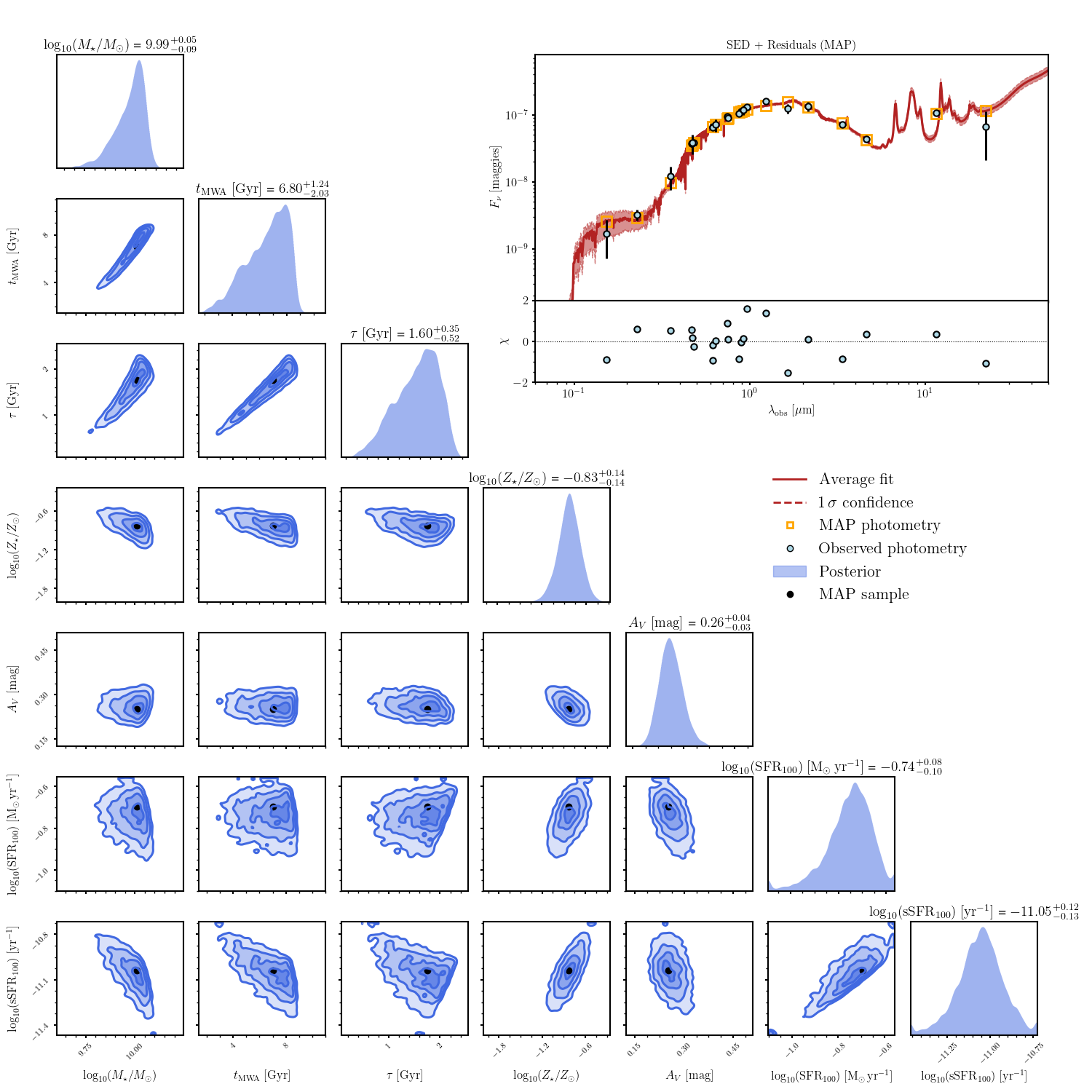}
    \caption{
        Results of our \texttt{Prospector} fit to the broadband SED of the host galaxy of \SNName. The main panel presents the posterior distributions for key stellar population parameters, including surviving stellar mass ($M_*$\,[\msun]), mass-weighted age ($t_\text{MWA}$\,[Gyr]), star-formation timescale ($\tau$\,[Gyr]), metallicity ($Z$\,[$Z_\odot$]), dust attenuation ($A_V$\,[AB mag]), and both average star formation rate (SFR$_{100}$\,[$M_\odot \, \text{yr}^{-1}$]) and specific SFR (sSFR$_{100}$\,[$\text{yr}^{-1}$]) over the past 100\,Myr. The contours correspond to $1 \sigma$, $2 \sigma$ and $3 \sigma$ confidence intervals, and the quoted values represent the median value and $1 \sigma$ uncertainties. The inset panel shows the observed photometry (blue points), the maximum a posteriori (MAP) model photometry (orange squares), and the best-fitting model SED (red curve) with $1 \sigma$ confidence interval (shaded). Residuals relative to the MAP model are also shown.}
    \label{FIG: prospector_fit}
\end{figure*}

We then modelled the host galaxy spectral energy distribution (SED) using the \texttt{Prospector} package \citep{Prospector}. We assumed a Chabrier initial mass function \citep{Chabrier2003}, the Milky Way extinction law \citep{Cardelli1989}, and enable the inclusion of circumstellar dust emission from asymptotic giant branch stars. The adopted model is a parametric delayed-$\tau$ star formation history (SFH $\propto te^{-t/\tau}$), with free parameters for total mass formed ($M_F$), age of the galaxy ($t_{\text{age}}$), e-folding timescale ($\tau$), and metallicity ($Z$). Because $t_{\text{age}}$ represents only the onset of star formation, a more physically meaningful measure of stellar population age is the mass-weighted age ($t_\mathrm{MWA}$), defined as:
\begin{equation}
    t_\text{MWA} = t_\text{age} - \frac{\int_0^{t_\text{age}}t \cdot \text{SFH}(t) dt}{\int_0^{t_\text{age}}\text{SFH}(t) dt}
\end{equation}
\citep[see][]{Nugent2020}.
Similarly, the relevant mass to report is the surviving stellar mass ($M_\star$); i.e., the present-day mass retained in stars and stellar remnants, rather than the total mass formed. This accounts for stellar mass loss over time, and can be approximated as:
\begin{equation}
    M_* \approx M_F \cdot 10^{(1.06 - 0.24\log_{10}[t_{\text{MWA}}  ({\rm yr})] + 0.01\log^2_{10}[t_{\text{MWA}} ({\rm yr})])}
\end{equation}
\citep[see][]{Leja2013}.
Sampling was performed using the \texttt{dynesty} nested sampler \citep{Dynesty}, providing robust posterior distributions. Stellar population synthesis models are constructed using \texttt{fsps} and \texttt{Python-fsps} \citep{Conroy2009, Conroy2010}. The resulting best-fit SED, along with photometry and residuals, is shown in Figure~\ref{FIG: prospector_fit}.

From this modelling, we infer the following host galaxy properties:
\begin{itemize}
    \item Stellar mass: $\log_{10}(M_*/{\rm M}_\odot)=9.99^{+0.05}_{-0.09}$
    \item Mass-weighted age: $t_\text{MWA}\,({\rm Gyr}) = 6.80^{+1.24}_{-2.03}$
    \item Star formation timescale: $\tau\,({\rm Gyr}) = 1.60^{+0.35}_{-0.52}$
    \item Metallicity: $\log_{10}(Z/Z_\odot)=-0.83^{+0.14}_{-0.14}$
    \item Dust extinction: $A_V = 0.26^{+0.04}_{-0.03}$
    \item Star formation rate (averaged over the past 100\,Myr): $\log_{10}(\text{SFR}_{100} \ [{\rm M}_\odot\,{\rm yr}^{-1}])=-0.74^{+0.08}_{-0.10}$
\end{itemize}

Next we compare these values to the host galaxy of \gfo. \GWxx{170817}/\gfo\ occurred in the sd0 galaxy NGC\,4993 \citep{Hjorth2017, Levan2017}. \cite{Stevance2023} estimated a stellar mass, \mbox{$\log_{10} (M_* / \mathrm{M}_{\odot}) = 10.4$} and a dominant stellar population older than 5\,Gyr (peaking around $7 - 12.5$\,Gyr) with metallicity, $Z = 0.010$. A younger stellar component (1\,Gyr) with enhanced metal content ($Z = 0.020 - 0.030$) was found to account for roughly 5\% of the total mass of NGC\,4993. From our host galaxy modelling of \HostGalName, we find similar galaxy ages and masses to those of NGC\,4993, but a substantially lower metallicity ($\sim 15$ vs.\ 70\% Solar metallicity).

\SNName\ is located 0.88'' from its host galaxy, which, at a distance of \HostDist\ (see Section~\ref{SEC: SN2025ulz - Redshift estimation}), corresponds to a projected physical offset of 1.45\,kpc. This is comparable to \gfo, which had a projected distance from NGC\,4993 of 1.96\,kpc \citep{Levan2017}.

\cite{Schulze2021} present an overview of SN host galaxy properties, including those of SNe~IIb. They measure median (mode) galaxy masses, $\log_{10} (M_* / {\rm \msun}) = 9.53_{-0.16}^{+0.15} \ (10.04_{-0.42}^{+0.15})$, ages of $2.2 \pm 0.3 \ (5.5_{-0.7}^{+0.8})$\,Gyr, star formation rates, $\log_{10}({\rm SFR} \ [{\rm M}_\odot\,{\rm yr}^{-1}])= -0.30 \pm 0.12 \ (-0.18_{-0.29}^{+0.25})$, and offsets of $2.70 \pm 0.45 \ (4.06 \pm 0.44)$\,kpc. \cite{Kelly2012} present metallicity measurements for core-collapse SN host galaxies; they measure a median host galaxy metallicity for SNe~IIb of $Z \approx 0.02$. \HostGalName\ matches the modal values of host galaxy mass and age, while possessing a lower-than-average metallicity and star formation rate. We note, however, that metallicity estimates derived from SED fitting to broadband photometry alone are highly uncertain and should be interpreted with caution, as they are generally not well constrained without spectroscopic information. Additionally, the host galaxy offset appears to be lower than average for SNe~IIb.

\section{Limits on KN emission from Pan-STARRS observations} \label{SEC: KN limits from Pan-STARRS}

\begin{figure*}
    \centering
    \includegraphics[width=\linewidth]{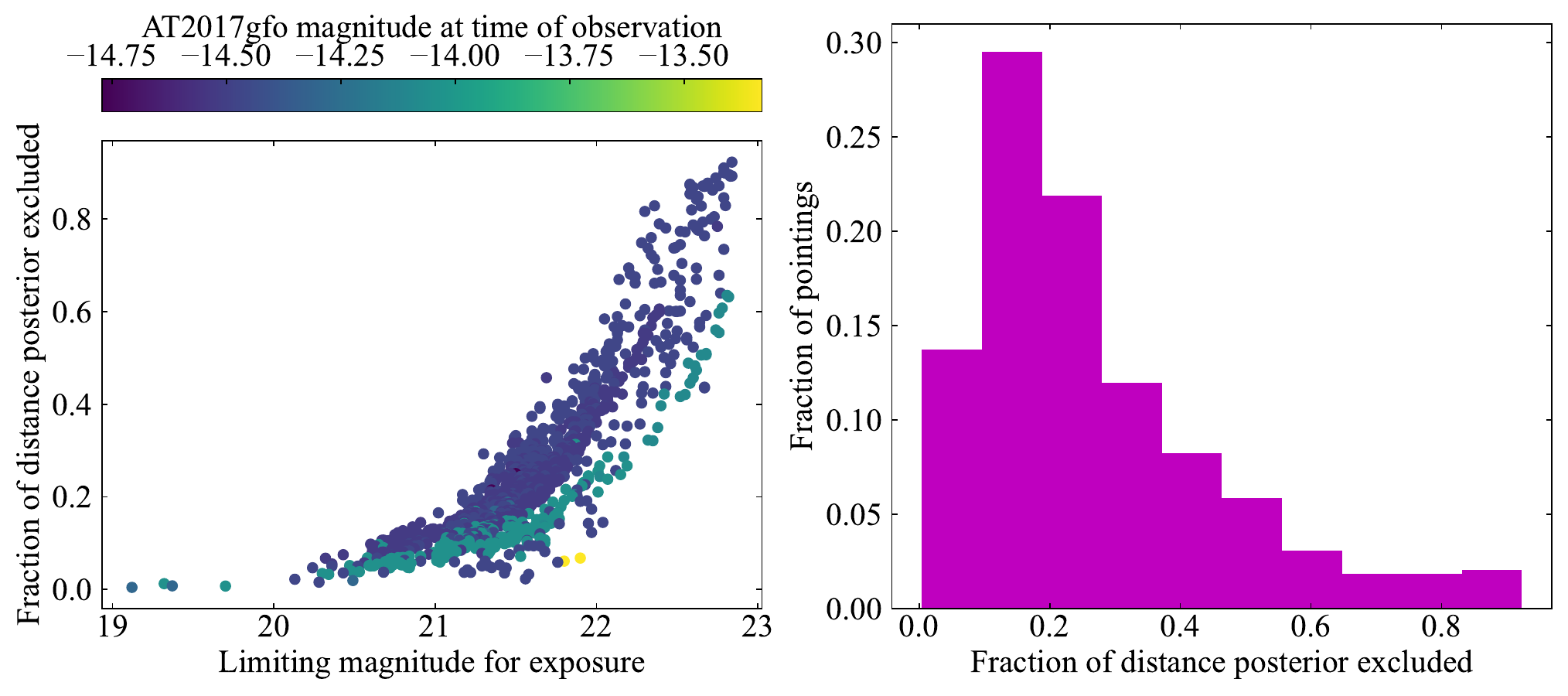}
    \caption{
        \textit{Left:} Illustration of the limiting magnitude reached by each Pan-STARRS pointing, and the corresponding fraction of the distance posterior for which \gfo-like KN emission can be ruled out. 
        \textit{Right:} Histogram highlighting the fraction of pointings that rule out different fractions of the distance posterior of \GWsource. Only a few of the pointings enable the ruling out of a substantial fraction ($\gtrsim 0.8$) of the distance posterior.
    }
    \label{FIG: KN limits}
\end{figure*}

With our multi-night pointings from Pan-STARRS, we do not identify any convincing candidate counterparts to \GWsource. Next, we investigate whether our observations can constrain the presence of KN emission within the region tiled by Pan-STARRS.

For every Pan-STARRS pointing, we extract an observation time relative to the GW trigger, and the $3.5 \sigma$ limiting magnitude. We then extract the absolute magnitude of the KN model for \gfo\ from \cite{Nicholl2021} at the same time, and in the same filter. This absolute magnitude estimate for \gfo\ was used to compute a distance at which it would have the same value as the limiting magnitude of each pointing (accounting for line-of-sight extinction). With this distance, we compute the fraction of the distance posterior of \GWsource\ from the \whichMap\ skymap that can be ruled out along that line of sight; i.e., we integrate the posterior over all distances closer than the distance computed for \gfo\ to remain at least as bright as our limiting magnitude.

The results of this analysis are plotted in Figure~\ref{FIG: KN limits}. Here the fraction of posterior distance excluded is plotted against the limiting magnitude of each pointing. From this plot, we can glean a number of useful insights. First, some of the Pan-STARRS pointings reach limiting magnitudes of $\lesssim 22.5$, which allow us to rule out a \gfo-like signal across $\gtrsim 80$\% of the distance posterior. Second, there is a spread in the fraction of the distance posterior excluded for pointings with similar limiting magnitudes. This can be explained by some combination of differences in distance posterior along different lines of sight, differences in extinction, and the expected magnitude at the time of observation. The data are colour-coded to indicate what magnitude \gfo\ would possess at the same phase as the Pan-STARRS pointings; \ie, the distributions of blue, turquoise and yellow points essentially correspond to \gfo\ at $\sim 2.2$, $\sim 3.2$ and $\sim 4.2$\,d post-burst. From this, one can see that the pointings from 20 August 2025 (MJD~60907.3; $\sim 2.2$\,d post-burst) are substantially more constraining than those from even just one day later. This visually illustrates how important rapid follow-up of GW localisations can be.

Also in Figure~\ref{FIG: KN limits} we present a histogram that visualises the fraction of pointings that rule out different fractions of distance posterior. A few of the pointings are quite discriminatory, ruling out $> 60$\% of the distance posterior. However, the bulk of our pointings are much less constraining, with the most common result being that we rule out just $10 - 20$\% of the distance posterior. 

With the above information, we also compute the fraction of the total surveyed volume within which we can confidently rule out the presence of a \gfo-like signal; we estimate this to be $\approx 27$\%. Considering the full \whichMap\ sky volume, this coverage equates to $\approx 9.1$\% of the total 90\% sky localisation volume. However, we note that if our observations were undertaken when the KN had $M_i = -15.7$ (the brightness of \gfo\ at +12\,hrs), we would be able to rule out the presence of a KN signal across 82\% of our surveyed volume. This substantial increase again highlights the paramount importance of expedient follow-up observations.

\section{Summary and Conclusions} \label{SEC: Summary and conclusions}

Here we have presented our observational campaign following the reported sub-threshold GW event, \GWsource. After establishing that many of the GW signal properties agree well with astrophysically real GW signals, we outline our observations tiling the \whichMap\ skymap with Pan-STARRS and ATLAS. Despite discovering 47 new extragalactic transient phenomena, further observations and analysis ruled all were unrelated to \GWsource. 

We next presented our follow-up observations of \SNName, a reported candidate counterpart to \GWsource. We measured a host galaxy redshift, $z =$~\Hostz, which places \SNName\ (just) within the $2 \sigma$ LVK distance estimate. Although the initial observations ($\lesssim 4$~days) mirror a KN-like decline, the lightcurve re-brightens after 5~days, more akin to a SN. We compare the full lightcurve of \SNName\ to a sample of SNe IIb, and find good agreement in early evolution. We also compare to \gfo\ and KN models, and determine that after $\lesssim 5$~days, the data are incompatible with a KN interpretation.

Finally, we explore how much of the \GWsource\ \whichMap\ sky volume we can exclude possessing a \gfo-like transient from our Pan-STARRS follow-up observations. We can exclude $\approx 27$\% of the distance posterior covered by Pan-STARRS, or $\approx 9.1$\% of the total \whichMap\ 90\% sky localisation volume. Were our observations executed more quickly (+12\,hrs, when \gfo\ possessed $M_i = -15.7$), we would exclude a \gfo-like transient within 82\% of the distance posterior covered by Pan-STARRS. Clearly, rapid follow-up observations are key to providing the best chance of detecting KN emission in future GW sky searches.


\movetabledown=5cm
\begin{rotatetable*}
\begin{deluxetable*}{lcccccccc}
\tablecaption{
    Pan-STARRS candidate counterpart discoveries.
    \label{TAB: Candidate list}
    }
\tabletypesize{\tiny}
\startdata
Pan-STARRS    &IAU TNS        &RA             &Dec.           &Discovery    &Discovery AB               &Discovery    &Redshift                                                     &Comments                                                \\
name          &name           &               &               &MJD          &magnitude                  &filter       &($z$)                                                        &                                                        \\
\midrule
PS25grz       &AT\,2025uso    &16:16:57.43    &+35:03:21.0    &60907.361    &$20.47 \pm 0.18$           &$i$          &$0.025 ^s$ (DESI DR1)                                        &Outside final \whichMap\ skymap. SN LC in ATLAS.        \\
PS25gsa       &AT\,2025usy    &17:55:12.50    &+56:02:02.0    &60907.375    &$20.95 \pm 0.27$           &$i$          &$0.090 \pm 0.012 ^p$ (LS)                                    &Flat LC. Beyond $2 \sigma$ LVK limit.                   \\
PS25gsh       &AT\,2025utr    &18:07:46.73    &+56:52:12.8    &60907.417    &$19.84 \pm 0.13$           &$i$          &$0.043 \pm 0.011 ^p$ (LS)                                    &SN LC in ATLAS.                                         \\
PS25gsm       &AT\,2025utx    &16:44:10.88    &+42:27:55.1    &60907.421    &$20.21 \pm 0.26$           &$i$          &$0.080 ^s$ (DESI DR1)                                        &Flat LC. Too bright ($M_i \sim -17.6$).                 \\
PS25gsv       &AT\,2025uuf    &20:40:40.26    &+64:36:41.4    &60907.377    &$20.99 \pm 0.29$           &$i$          &$0.061 ^p$ (2MASS)                                           &Flat LC.                                                \\
PS25gtm       &AT\,2025uuw    &16:36:55.61    &+44:22:48.1    &60907.373    &$20.44 \pm 0.20$           &$i$          &$0.034 \pm 0.006 ^p$ (LS)                                    &Classified as a SN Ia.                                  \\
PS25guo       &AT\,2025uxs    &15:44:14.34    &+27:28:41.9    &60909.291    &$20.05 \pm 0.08$           &$i$          &$0.063 ^s$ (DESI DR1)                                        &SN LC in ATLAS.                                         \\
\midrule
PS25grv       &AT\,2025usk    &16:04:49.18    &+35:47:26.8    &60907.355    &$20.81 \pm 0.29$           &$i$          &$0.229 \pm 0.039 ^p$ (LS)                                    &Too distant.                                            \\
PS25grw       &AT\,2025usl    &15:48:34.68    &+32:17:37.6    &60907.300    &$21.57 \pm 0.24$           &$g$          &$0.218 \pm 0.282 ^p$ (LS)                                    &Likely a CV. Alternatively,                             \\
"             &"              &"              &"              &"            &"                          &"            &$0.050 ^s$ (SDSS)                                            &offset from galaxy at $z = 0.05$, but flat LC.          \\
PS25grx       &AT\,2025usm    &15:47:43.01    &+30:32:53.5    &60907.309    &$21.69 \pm 0.25$           &$r$          &$0.319 \pm 0.069 ^p$ (LS)                                    &Too distant.                                            \\
PS25gry       &AT\,2025usn    &15:50:27.14    &+30:20:01.2    &60907.298    &$21.05 \pm 0.15$           &$g$          &$0.217 ^s$ (DESI DR1)                                        &Too distant.                                            \\
PS25gsg       &AT\,2025utq    &17:31:46.85    &+54:47:47.6    &60907.374    &$20.12 \pm 0.15$           &$i$          &$0.075 \pm 0.011 ^p$ (LS)                                    &Flat LC. Too bright ($M_i \sim -18$).                   \\
PS25gsi       &AT\,2025utt    &17:50:21.38    &+54:57:56.2    &60907.375    &$20.34 \pm 0.20$           &$i$          &$0.269 ^s$ (DESI DR1)                                        &Too distant.                                            \\
PS25gsj       &AT\,2025utu    &20:07:56.23    &+61:20:47.4    &60907.412    &$20.81 \pm 0.16$           &$i$          &$0.169 \pm 0.020 ^p$ (SDSS)                                  &Too distant.                                            \\
PS25gsk       &AT\,2025utv    &17:15:00.36    &+51:22:01.3    &60907.373    &$20.01 \pm 0.14$           &$i$          &$0.105 \pm 0.068 ^p$ (LS)                                    &Flat LC.                                                \\
PS25gsl       &AT\,2025utw    &19:47:51.84    &+64:30:33.0    &60907.425    &$20.59 \pm 0.23$           &$i$          &$0.899 \pm 0.544 ^p$ (LS)                                    &Too distant.                                            \\
PS25gsn       &AT\,2025uty    &17:41:51.38    &+52:25:50.3    &60907.408    &$20.74 \pm 0.23$           &$i$          &$0.646 \pm 0.204 ^p$ (LS)                                    &Too distant.                                            \\
PS25gso       &AT\,2025utz    &16:31:07.87    &+40:42:44.0    &60907.357    &$20.70 \pm 0.30$           &$i$          &$0.279 \pm 0.032 ^p$ (LS)                                    &Too distant.                                            \\
PS25gsp       &AT\,2025uua    &17:25:45.69    &+51:55:43.2    &60907.373    &$20.87 \pm 0.26$           &$i$          &$0.280 \pm 0.077 ^p$ (LS)                                    &Too distant.                                            \\					
PS25gss       &AT\,2025uub    &20:01:03.14    &+61:58:53.8    &60907.396    &$19.16 \pm 0.06$           &$i$          &---                                                          &High proper motion star.                                \\					
PS25gst       &AT\,2025uuc    &17:39:58.54    &+53:20:15.8    &60907.411    &$20.89 \pm 0.31$           &$i$          &$0.305 \pm 0.109 / 0.158 \pm 0.201 ^p$ (LS)                  &Either too distant, or a likely CV.                     \\
PS25gsu       &AT\,2025uud    &20:53:53.82    &+65:10:57.2    &60907.450    &$20.06 \pm 0.12$           &$i$          &---                                                          &Flat LC. Possibly Galactic.                             \\					
PS25gsx       &AT\,2025uug    &19:02:03.80    &+59:19:37.5    &60907.376    &$20.64 \pm 0.15$           &$i$          &$0.145 \pm 0.022 ^p$ (LS)                                    &Too distant.                                            \\					
PS25gsy       &AT\,2025uuh    &18:54:47.78    &+60:12:28.0    &60907.488    &$21.48 \pm 0.28$           &$i$          &$0.456 \pm 0.459 ^p$ (LS)                                    &Flat LC. Very uncertain redshift.                       \\
PS25gsz       &AT\,2025uui    &15:58:55.23    &+30:59:35.9    &60907.304    &$20.86 \pm 0.19$           &$g$          &$0.174 ^s$ (DESI DR1)                                        &Too distant.                                            \\					
PS25gta       &AT\,2025uuk    &17:29:35.00    &+54:36:16.3    &60907.374    &$20.30 \pm 0.24$           &$i$          &$0.202 \pm 0.006 ^p$ (LS)                                    &Too distant.                                            \\					
PS25gtc       &AT\,2025uuo    &17:11:33.53    &+51:33:37.2    &60907.373    &$20.04 \pm 0.19$           &$i$          &$0.128 \pm 0.008 ^p$ (LS)                                    &Too distant.                                            \\					
PS25gtd       &AT\,2025uup    &19:50:58.53    &+61:33:04.7    &60907.396    &$20.96 \pm 0.21$           &$i$          &$0.411 \pm 0.121 ^p$ (LS)                                    &Too distant.                                            \\					
PS25gte       &AT\,2025uuq    &16:18:35.31    &+37:56:33.2    &60907.363    &$20.21 \pm 0.24$           &$i$          &$0.148 \pm 0.028 ^p$ (LS)                                    &Too distant.                                            \\					
PS25gtf       &AT\,2025uur    &15:47:00.15    &+30:31:38.2    &60907.298    &$21.87 \pm 0.35$           &$g$          &$0.294 \pm 0.111 ^p$ (LS)                                    &Too distant.                                            \\					
PS25gtg       &AT\,2025uus    &16:17:10.71    &+39:38:55.0    &60907.357    &$20.83 \pm 0.27$           &$i$          &$0.142 \pm 0.018 ^p$ (LS)                                    &Too distant.                                            \\					
PS25gth       &AT\,2025uut    &18:54:16.71    &+60:01:57.4    &60907.396    &$20.81 \pm 0.22$           &$i$          &$0.093 \pm 0.030 ^p$ (LS)                                    &Flat LC.                                                \\					
PS25gti       &AT\,2025uuu    &17:48:26.47    &+56:00:47.8    &60907.375    &$21.32 \pm 0.29$           &$i$          &$0.374 \pm 0.088 ^p$ (LS)                                    &Too distant.                                            \\					
PS25gtj       &AT\,2025uuv    &20:46:53.44    &+65:07:37.0    &60907.432    &$20.67 \pm 0.27$           &$i$          &---                                                          &No redshift. Flat LC out to MJD~60924.3.                \\
PS25gtt       &AT\,2025uvs    &17:56:55.28    &+55:21:05.5    &60907.375    &$21.24 \pm 0.27$           &$i$          &$1.211 ^s$ (DESI DR1)                                        &Too distant.                                            \\					
PS25gtu       &AT\,2025uvt    &15:55:32.44    &+31:00:35.1    &60907.323    &$22.16 \pm 0.32$           &$i$          &$2.710 ^s$ (DESI DR1)                                        &Too distant.                                            \\					
PS25gtv       &AT\,2025uvu    &15:46:51.46    &+29:59:58.0    &60907.298    &$21.94 \pm 0.29$           &$g$          &$0.784 ^s$ (DESI DR1)                                        &Too distant.                                            \\					
"             &"              &"              &"              &"            &"                          &"            &Nearby galaxies have $z = 0.112, 0.113 ^s$ (DESI DR1)        &Also too distant.                                       \\					
PS25gum       &AT\,2025uxm    &15:56:59.45    &+29:25:24.5    &60907.306    &$21.82 \pm 0.25$           &$g$          &$1.350 ^s$ (DESI DR1)                                        &Too distant.                                            \\				
PS25gun       &AT\,2025uxo    &16:35:33.20    &+41:00:57.9    &60909.301    &$20.57 \pm 0.17$           &$i$          &$0.127 \pm 0.036 ^p$ (LS)                                    &Too distant.                                            \\	
PS25guq       &AT\,2025uya    &16:34:33.59    &+43:57:01.3    &60909.272    &$20.80 \pm 0.17$           &$i$          &$0.272 \pm 0.033 ^p$ (LS)                                    &Too distant.                                            \\					
PS25hah       &AT\,2025vjj    &15:45:16.27    &+29:31:58.6    &60912.251    &$22.05 \pm 0.29$           &$i$          &$0.112 ^s$ (DESI DR1)                                        &Too distant.                                            \\					
PS25hcu       &AT\,2025war    &19:53:44.05    &+60:42:10.6    &60907.412    &$20.44 \pm 0.11$           &$i$          &$0.144 \pm 0.028 ^p$ (SDSS)                                  &Too distant.                                            \\					
PS25hcv       &AT\,2025wat    &16:01:46.84    &+31:13:29.9    &60913.257    &$22.38 \pm 0.23$           &$i$          &$0.419 \pm 0.053 ^p$ (LS)                                    &Too distant.                                            \\					
"             &"              &"              &"              &"            &"                          &"            &Nearby galaxies have $z = 0.072, 0.030 ^s$ (DESI DR1)        &SN-like LC evolution.                                   \\
PS25hdv       &AT\,2025wek    &15:48:19.75    &+32:45:36.8    &60907.302    &$21.92 \pm 0.30$           &$g$          &$0.741 ^s$ (DESI DR1)                                        &Too distant.                                            \\					
PS25hdw       &AT\,2025wel    &15:42:57.22    &+30:24:56.9    &60909.264    &$22.05 \pm 0.30$           &$i$          &$0.449 \pm 0.067 ^p$ (LS)                                    &Too distant.                                            \\					
PS25hdx       &AT\,2025wfr    &15:50:30.57    &+29:21:17.7    &60913.250    &$21.98 \pm 0.20$           &$i$          &$0.387 \pm 0.086 ^p$ (LS)                                    &Too distant.                                            \\					
PS25hdy       &AT\,2025wfs    &15:46:17.15    &+31:50:19.8    &60910.263    &$22.22 \pm 0.30$           &$i$          &$0.266 \pm 0.086 ^p$ (LS)                                    &Too distant.                                            \\		
\enddata
\tablecomments{
    \\
    LC~=~lightcurve.
    \\
    $^{s} =$~spectroscopically derived redshift;
    $^{p} =$~photometrically derived redshift.
    \\
    References:
        LS = \cite{LegacySurveys_DR9_DataRelease};
        SDSS = \cite{SDSSDR15};
        DESI DR1 = \cite{DESI_DR1_DataRelease};
        2MASS = \cite{Bilicki2014_2MASS}.
}
\end{deluxetable*}
\end{rotatetable*}

\startlongtable
\begin{deluxetable*}{lccrc}
\tabletypesize{\tiny}
\tablewidth{0pt}
\tablecaption{
    Observations of \SNName. Here $t_0$ denotes the time reported for the GW event \citep[MJD \MJDGW;][]{LVKGCNDiscovery}. Magnitudes are not corrected for the expected foreground extinction of \mbox{$E (B - V) =$~\EBsubV} \citep{Schlafly2011}. Limiting magnitudes are quoted to $3 \sigma$ significance.
}
\label{TAB: SN2025ulz photometry}
\tablehead{
    \colhead{$t - t_{0}$} & \colhead{MJD} & \colhead{Filter} & \colhead{Total exposure} & \colhead{Apparent magnitude} \\
    [-1.5em]
    \colhead{(days)} & \colhead{} & \colhead{} & \colhead{time (s)} & \colhead{(AB mag)}
}
\startdata
    \multicolumn{5}{l}{Pan-STARRS}    \\
    \midrule
    $-9.737$    &60895.319    &$i$    &180     &$> 21.00$             \\
    $-0.823$    &60904.233    &$y$    &120     &$> 19.70$             \\
    2.270       &60907.325    &$i$    &900     &$22.27 \pm 0.24$      \\
    2.273       &60907.329    &$r$    &1800    &$22.24 \pm 0.12$      \\
    2.281       &60907.337    &$z$    &900     &$21.88 \pm 0.22$      \\
    2.290       &60907.345    &$g$    &1800    &$22.85 \pm 0.29$      \\
    2.307       &60907.362    &$y$    &1800    &$> 21.40$             \\
    3.193       &60908.248    &$y$    &1800    &$> 19.62$             \\
    3.198       &60908.254    &$z$    &1800    &$> 20.70$             \\
    3.215       &60908.270    &$g$    &1800    &$> 21.53$             \\
    3.218       &60908.273    &$i$    &1350    &$> 21.09$             \\
    4.202       &60909.257    &$r$    &1200    &$22.42 \pm 0.29$      \\
    4.203       &60909.258    &$i$    &1800    &$22.52 \pm 0.32$      \\
    4.216       &60909.272    &$g$    &1200    &$> 22.91$             \\
    5.206       &60910.262    &$r$    &1800    &$22.54 \pm 0.24$      \\
    5.208       &60910.263    &$i$    &1800    &$22.54 \pm 0.26$      \\
    5.223       &60910.279    &$g$    &1800    &$> 22.62$             \\
    6.201       &60911.257    &$i$    &1800    &$22.15 \pm 0.19$      \\
    6.217       &60911.272    &$r$    &1800    &$22.07 \pm 0.14$      \\
    6.223       &60911.279    &$g$    &1800    &$> 22.65$             \\
    7.205       &60912.260    &$r$    &1800    &$21.95 \pm 0.10$      \\
    7.208       &60912.263    &$i$    &1800    &$21.63 \pm 0.12$      \\
    7.222       &60912.278    &$g$    &1800    &$22.58 \pm 0.29$      \\
    8.200       &60913.255    &$i$    &1800    &$21.50 \pm 0.07$      \\
    8.216       &60913.271    &$r$    &1800    &$21.58 \pm 0.06$      \\
    8.223       &60913.278    &$g$    &1800    &$22.28 \pm 0.15$      \\
    10.200      &60915.256    &$i$    &1800    &$21.41 \pm 0.12$      \\
    10.201      &60915.257    &$r$    &1800    &$21.55 \pm 0.11$      \\
    10.223      &60915.278    &$g$    &1800    &$22.12 \pm 0.21$      \\
    12.200      &60917.256    &$i$    &1800    &$21.07 \pm 0.07$      \\
    12.210      &60917.266    &$r$    &1800    &$21.22 \pm 0.08$      \\
    12.222      &60917.278    &$g$    &1800    &$21.81 \pm 0.18$      \\
    15.201      &60920.257    &$i$    &1800    &$20.83 \pm 0.07$      \\
    15.216      &60920.272    &$r$    &1800    &$21.05 \pm 0.11$      \\
    15.223      &60920.278    &$g$    &1800    &$21.48 \pm 0.17$      \\
    17.192      &60922.248    &$i$    &1800    &$20.66 \pm 0.06$      \\
    17.215      &60922.270    &$g$    &1800    &$21.52 \pm 0.22$      \\
    17.237      &60922.292    &$r$    &1800    &$21.26 \pm 0.29$      \\
    20.197      &60925.253    &$i$    &600     &$20.76 \pm 0.14$      \\
    20.205      &60925.261    &$g$    &600     &$> 20.37$             \\
    20.213      &60925.269    &$r$    &600     &$> 20.83$             \\
    23.197      &60928.253    &$i$    &600     &$20.98 \pm 0.15$      \\
    23.205      &60928.260    &$g$    &600     &$> 21.59$             \\
    23.213      &60928.268    &$r$    &600     &$21.11 \pm 0.24$      \\
    26.190      &60931.246    &$i$    &600     &$20.71 \pm 0.09$      \\
    26.198      &60931.254    &$g$    &600     &$22.07 \pm 0.25$      \\
    26.206      &60931.262    &$r$    &600     &$20.88 \pm 0.14$      \\
    29.184      &60934.240    &$i$    &600     &$21.13 \pm 0.19$      \\
    29.192      &60934.247    &$g$    &600     &$> 21.82$             \\
    29.199      &60934.255    &$r$    &600     &$21.38 \pm 0.24$      \\
    35.189      &60940.245    &$i$    &1800    &$21.56 \pm 0.19$      \\
    38.179      &60943.234    &$i$    &600     &$21.75 \pm 0.31$      \\
    38.187      &60943.242    &$r$    &600     &$> 21.87$             \\
    38.194      &60943.250    &$g$    &600     &$> 21.72$             \\
    42.171      &60947.227    &$i$    &600     &$> 21.33$             \\
    42.180      &60947.235    &$r$    &600     &$> 21.98$             \\
    42.188      &60947.243    &$g$    &600     &$> 21.85$             \\
    45.172      &60950.227    &$i$    &600     &$21.55 \pm 0.27$      \\
    45.180      &60950.235    &$r$    &600     &$> 21.52$             \\
    48.174      &60953.230    &$i$    &600     &$> 21.39$             \\
    48.182      &60953.238    &$r$    &600     &$> 21.05$             \\
    52.167      &60957.222    &$i$    &600     &$21.98 \pm 0.30$      \\
    52.174      &60957.230    &$r$    &600     &$> 21.94$             \\
    58.160      &60963.216    &$i$    &600     &$> 21.41$             \\
    58.169      &60963.225    &$r$    &600     &$> 21.37$             \\
    62.159      &60967.214    &$i$    &1200    &$> 20.91$             \\
    \midrule
    \multicolumn{5}{l}{ATLAS}    \\
    \midrule
    $-1.734$    &60903.322    &$o$    &440    &$> 21.60$              \\
    0.860       &60905.916    &$w$    &120    &$> 21.70$              \\
    \midrule
    \multicolumn{5}{l}{LOT}    \\
    \midrule
    5.459     &60910.515    &$g$    &900     &$> 22.35$               \\
    5.463     &60910.518    &$r$    &900     &$> 22.04$               \\
    5.531     &60910.587    &$u$    &1800    &$> 21.14$               \\
    11.526    &60916.582    &$g$    &1800    &$22.06 \pm 0.12$        \\
    11.547    &60916.602    &$r$    &1500    &$> 21.08$               \\
    12.507    &60917.563    &$g$    &1800    &$22.01 \pm 0.17$        \\
    18.511    &60923.567    &$g$    &1800    &$> 20.82$               \\
    26.419    &60931.475    &$g$    &1800    &$22.22 \pm 0.11$        \\
    26.441    &60931.497    &$r$    &1800    &$21.24 \pm 0.07$        \\
    \midrule
    \multicolumn{5}{l}{SLT}    \\
    \midrule
    10.493    &60915.548    &$i$    &7200    &$> 21.25$               \\
    11.487    &60916.542    &$i$    &2400    &$> 20.93$               \\
    28.445    &60933.501    &$i$    &3600    &$> 21.05$               \\
\enddata
\end{deluxetable*}

\section*{Data Availability}

The updated Pan-STARRS reference images utilised for accurate difference imaging of \SNName\ (see Section~\ref{SEC: SN2025ulz - Initial discovery and observations}) are publicly available at \url{https://ora.ox.ac.uk/objects/uuid:624c1bc5-b841-4da0-9c56-c8683454da7f}.

\section*{Acknowledgments}

SJS, SS, KS, DY, JHG, FS, JWT acknowledge funding from STFC grants ST/Y001605/1, ST/X001253/1, ST/Y509474/1, a Royal Society Research Professorship, a Royal Society Newton International Fellowship and the Hintze Family Charitable Foundation. 
T.-W.C. and A.A. acknowledge the financial support from the Yushan Fellow Program by the Ministry of Education, Taiwan (MOE-111-YSFMS-0008-001-P1) and the National Science and Technology Council, Taiwan (NSTC grant 114-2112-M-008-021-MY3).
This work was funded by ANID, Millennium Science Initiative, ICN12\_009.
Pan-STARRS is primarily funded to search for near-earth asteroids through NASA grants NNX08AR22G and NNX14AM74G. The Pan-STARRS science products for LIGO--Virgo--KAGRA follow-up are made possible through the contributions of the University of Hawaii's Institute for Astronomy and Queen's University Belfast.
Parts of this research were supported by the Australian Research Council Centre of Excellence for Gravitational Wave Discovery (OzGrav), through project number CE230100016.

The Pan-STARRS1 Sky Surveys have been made possible through contributions by the University of Hawaii's Institute for Astronomy, the Pan-STARRS Project Office, the Max Planck Society and its participating institutes, the Max Planck Institute for Astronomy, Heidelberg and the Max Planck Institute for Extraterrestrial Physics, Garching, Johns Hopkins University, Durham University, the University of Edinburgh, Queen's University Belfast, the Harvard-Smithsonian Center for Astrophysics, the Las Cumbres Observatory Global Telescope Network Incorporated, the National Central University of Taiwan, the Space Telescope Science Institute, and the National Aeronautics and Space Administration under Grant No.~NNX08AR22G issued through the Planetary Science Division of the NASA Science Mission Directorate, the National Science Foundation Grant No.~AST-1238877, the University of Maryland, Eotvos Lorand University, and the Los Alamos National Laboratory.
This work has made use of data from the Asteroid Terrestrial-impact Last Alert System project. ATLAS is primarily funded to search for near-earth asteroids (NEOs) through NASA grants NN12AR55G, 80NSSC18K0284 and 80NSSC18K1575; by-products of the NEO search include images and catalogs from the survey area.  The ATLAS science products have been made possible through the contributions of the University of Hawaii's Institute for Astronomy, Queen's University Belfast, the Space Telescope Science Institute and the South African Astronomical Observatory.
This publication makes use of data collected at Lulin Observatory, which is partly supported by the TAOVA grant NSTC 114-2740-M-008-002.

This research has made use of the NASA/IPAC Extragalactic Database (NED), which is operated by the Jet Propulsion Laboratory, California Institute of Technology, under contract with the National Aeronautics and Space Administration.

\facilities{
    \begin{itemize}
        \item Pan-STARRS,
        \item ATLAS,
        \item 1\,m LOT,
        \item 40\,cm SLT,
        \item UH 2.2\,m (SNIFS).
    \end{itemize}
}

\software{
    \begin{itemize}
        \item Astropy \citep{2013A&A...558A..33A, 2018AJ....156..123A, 2022ApJ...935..167A},
        \item Scipy \citep{2020SciPy-NMeth},
        \item MOSFiT \citep{guillochon2018mosfit},
        \item Photutils \citep{photutils}.
    \end{itemize}
}

\bibliography{References}
\bibliographystyle{aasjournal}

\end{document}